\documentclass[aip,reprint,pop]{revtex4-1}

\usepackage{graphicx}
\usepackage{subfigure}
\usepackage{epstopdf}
\usepackage{amsmath}
\usepackage{nicefrac}
\usepackage{dcolumn}
\usepackage{multirow}
\usepackage{bm}
\usepackage{mathrsfs}
\usepackage[colorlinks=true, linkcolor=blue]{hyperref}

\begin{document}

\title{$\mathbf{E}\times\mathbf{B}$ flows  for high-throughput plasma mass separation}
\author{Renaud Gueroult}
\affiliation{LAPLACE, Universit\'{e} de Toulouse, CNRS, 31062 Toulouse, France}
\author{Stewart J. Zweben}
\author{Nathaniel J. Fisch}
\affiliation{Princeton Plasma Physics Laboratory, Princeton University, Princeton, NJ 08543, USA}
\author{J.-M. Rax}
\affiliation{LOA-ENSTA, Universit\'{e} de Paris XI-Ecole Polytechnique, 91128 Palaiseau, France}

\begin{abstract}
High-throughput plasma separation based on atomic mass holds the promise for offering unique solutions to a variety of high-impact societal applications. Through the mass differential effects they exhibit, crossed-field configurations can in principle be exploited in various ways to separate ions based on atomic mass. Yet, the practicality of these concepts is conditioned upon the ability to drive suitable crossed-field flows for plasma parameters compatible with high-throughput operation. Limited current predictive capabilities have not yet made it possible to confirm this possibility. Yet, past experimental results suggest that end-electrodes biasing may be effective, at least for certain electric field values. A better understanding of cross-field conductivity is needed to confirm these results and confirm the potential of crossed-field configurations for high-throughput separation. 
\end{abstract}

\date{\today}
\maketitle

\section{Introduction}

Separation processes are critical steps in many industries. Yet, the efficiency of most industrial chemical separation processes, including the widely used distillation techniques, remains well below thermodynamic limits~\cite{Cussler2012}. Developing energy-efficient separation processes therefore holds significant upside potential both for energy and the environment. For instance, transitioning to improved separation processes is projected to reduce energy costs by $\$4$ billion per year in the U.S. petroleum, chemical and paper manufacturing industries alone~\cite{Sholl2016}. 

Compared to neutral particles in liquid and gases, electrically charged particles can in principle be manipulated in many more ways. One possibility is to use particles' electric charge to enhance the efficiency of existing separation processes. This is for example the idea behind electrofiltration, which takes advantage of the electric charge naturally present on certain particles to facilitate their separation through porous membranes~\cite{Jou2014,Kang2015}. A different approach consists in using electric and magnetic forces as the primary mechanism to separate particles. For this approach to be efficient, a large enough fraction of particles needs to be charged, which requires operating in an ionized gas or plasma.

The first and arguably best known separation technique utilizing charged particles is the mass spectrometer originally designed by Dempster~\cite{Dempster1918} and  Aston~\cite{Aston1919} following pioneering work by Wien~\cite{Wien1898} and Thomson~\cite{Thomson1913} (see, \emph{e.~g.}, Refs.~\cite{Beynon1978,Muenzenberg2013} for a historical account of the development of mass spectrometry). In these devices, charged particles are accelerated through a voltage gap and then separated based on charge to mass ratio $q/m$ in a region permeated by a perpendicular magnetic field. This technique forms the basis for separation based on differences in gyro-orbit, and was implemented and used at large scale in calutrons~\cite{Lawrence1958} for isotope separation during the second world war. Yet, separation in these devices relies on single particle motion which sets constrains on the practical operating parameter space. Instabilities~\cite{Alexeff1978} and space charge effects~\cite{Smith1947,Parkins2005} are known to impede high-density operations, which in turn limits practical throughput in these devices.

While high-throughput separation is always a desirable property, the acute need for it has only emerged within the last decade. Indeed, up until 2000s, plasma separation had mainly been considered for isotope separation~\cite{Boeschoten1979,Louvet1989,Grossman1991,Dolgolenko2017}, where quantities to be separated are typically a few tens of kg yr$^{-1}$. However, it has since then been recognized that plasma separation, and more specifically plasma separation based on atomic mass, could offer unique solutions to outstanding societal challenges including nuclear waste cleanup~\cite{Gueroult2015}, nuclear spent fuel reprocessing~\cite{Timofeev2014,Gueroult2014a,Vorona2015,Dolgolenko2017,Yuferov2017} and rare earth elements recycling~\cite{Gueroult2018a}. In addition to promising efficient solutions to separation needs which are particularly challenging for conventional chemical techniques, an important advantage of plasma separation for these application is that it is anticipated to have a much smaller environmental footprint. Yet, in contrast with isotope separation, these applications typically involve processing many tens of tons per year. In addition, these new applications differ from isotope separation in that the mass difference between elements to be separated is typically a few tens of atomic mass units (see Ref.~\cite{Gueroult2018} for an in-depth comparative analysis of the separation needs of these three applications). For plasma separation to be practical for these applications, plasma separation devices capable of throughputs of $10^4$ kg yr$^{-1}$ with large mass differences are therefore called for.

To address this new need, various concepts relying on different physical phenomena have been proposed~\cite{Dolgolenko2017}. Yet, high-throughput plasma separation poses its own set of physics and technological challenges, as recently reviewed by Zweben~\emph{et al.}~\cite{Zweben2018}. Identifying and addressing these challenges is the first step towards the development of practical devices. One challenge common to many concepts is the need for driving flows in crossed-field configurations.

This paper is organized as follows. In Sec.~\ref{Sec:SecII}, we begin by reviewing the various crossed-field plasma mass separation concepts proposed to date, along with their particular cross-field flow control requirements. In Sec.~\ref{Sec:SecIII}, the basic physics picture for driving cross-field flow using electrodes along magnetic field lines is presented, and past experimental results providing insights into the practicality of this scheme for plasma separation are discussed. Finally, the main findings are summarized in Sec.~\ref{Sec:SecIV}.


\section{Mass separation in crossed-field configurations}
\label{Sec:SecII}

Many different plasma separation schemes have been proposed over the years. In this paper, we focus on one particular family of plasma filtering concepts, namely plasma separation concepts relying on crossed-field configurations, with the goal of highlighting the challenges towards the realization of these concepts. Broader surveys and discussion of plasma filter concepts, including those based ion-cyclotron resonance~\cite{Dawson1976}, drift in curved magnetic field~\cite{Timofeev2000} and collisionality gradients~\cite{Ochs2017a} can be found in recent reviews~\cite{Dolgolenko2017,Zweben2018}. 

Crossed-field separation schemes can be broadly divided into two groups depending on whether ions are magnetized or not, that is to say whether $r_{ci}/L$ is smaller or larger than $1$. Here $L$ is the device characteristic length across the magentic field and $r_{ci}$ is the gyro-radius of an ion with thermal speed. This characteristic is used in this section to introduce various plasma filter configurations proposed to date.

\subsection{Crossed-field in magnetized ion regime}
\label{Sec:SecIIa}

Fluids in rotating motion experience a density dependent centrifugal force. In neutral fluids (liquids and gases), fluid rotation must be imparted mechanically. This is generally achieved through entrainement at the edge by moving parts (\emph{e.~g.} a rotor). Bulk rotation then results from viscous forces between fluid elements. In steady state, a pressure gradient forms to balance out the radially outward flux induced by centrifugal forces. Since the equilibrium profile depends on the molecule mass and the rotation frequency $\Omega$, the composition of a mixture will vary with radius, enabling separation~\cite{Soubbaramayer1979}. Yet, a limit of neutral centrifuges is that the rotation velocity $\Omega r$, which affects directly the separating power, is constrained by the mechanical stress exerted on moving parts. Since rotation in plasmas can in principle be produced in volume by taking advantage of particles electric charge, rotating plasmas can in principle address this shortcoming and access higher rotation velocities.

\begin{figure*}[htbp]
\begin{center}
\includegraphics[]{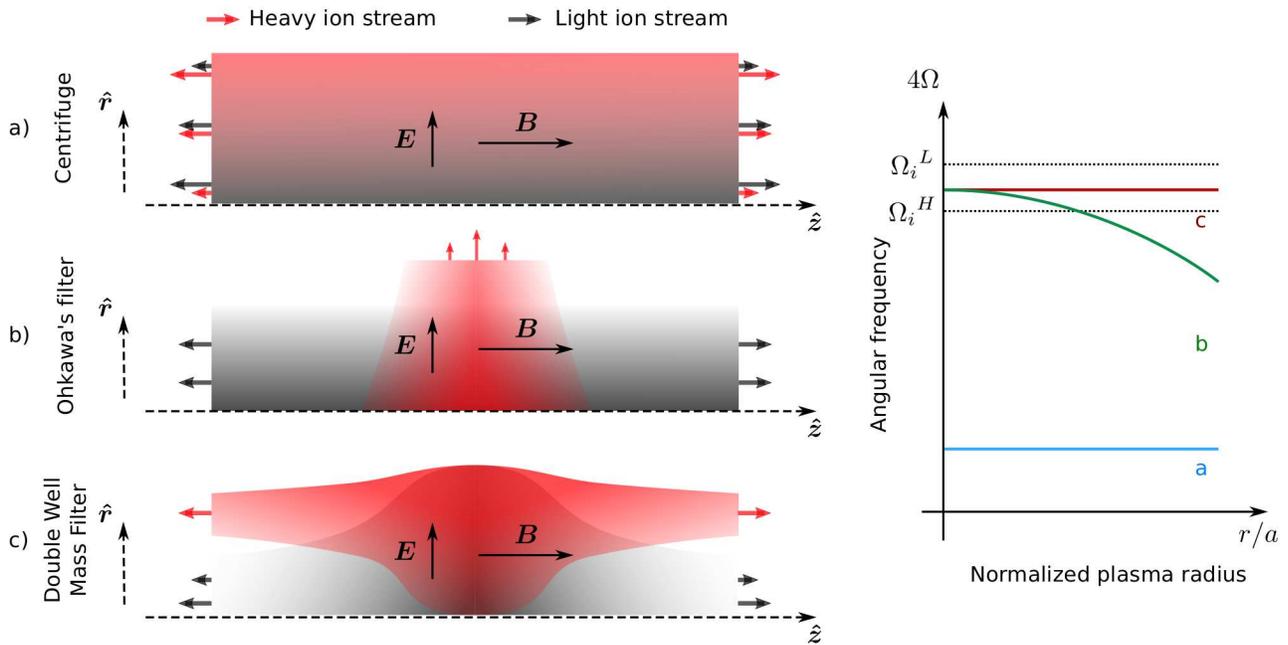}
\caption{Sketch of the separation process and required rotation frequency $\Omega$ for (a) a plasma centrifuge~\cite{Bonnevier1966}, (b) Ohkawa's filter~\cite{Ohkawa2002} and (c) the Double Well Mass Filter~\cite{Gueroult2014}. Although all three concepts have the same generic crossed-field configuration [$\bm{E} = E\bm{\hat{r}}$,$\bm{B} = B_0\bm{\hat{z}}$] and operate in magnetized ion regime, the different radial potential profile required by each concept translates into very different separation flows. ${\Omega_i}^H$ and ${\Omega_i}^L$ designate the heavy and light ion gyro-frequency, respectively. $a$ is the plasma column radius. Thick red and grey arrows represent heavy and light ion flows, respectively. The longer the arrow, the larger the flow.}
\label{Fig:Crossed_field_magnetized}
\end{center}
\end{figure*}

Since Bonnevier's original calculation that mass differential effects should arise from diffusion in a multi-ion species plasma subjected to centrifugal forces~\cite{Bonnevier1966}, crossed-field or $\bm{E}\times\bm{B}$ configurations have played a central role in separation in rotating magnetized plasma~\cite{Lehnert1971}. Here $\bm{E}$ and $\bm{B}$ are the electric and magnetic fields, respectively. In uniform fields, ion and electron crossed-field drifts are equal, and no mass or charge dependent motion arise. On the other hand, in non-uniform crossed-field configurations such as a radial electric field $\bm{E} = E\bm{\hat{r}}$ in a cylindrical plasma column permeated by a uniform axial magnetic field $\bm{B} = B_0\bm{\hat{z}}$, inertial terms introduce charge and mass asymmetries. In particular, ions will exhibit mass dependent azimuthal drift velocities. The azimuthal collisional drag force exerted by the slower light ions on the faster heavy ions, and vice-versa, in turn leads to an inward drift of light ions and an outward drift of heavy ions. In steady-state, heavy ions are then found preferentially at larger radius, while light ions are preferentially found at smaller radius~\cite{Oneil1981,Bittencourt1987}. This is illustrated in Fig.~\ref{Fig:Crossed_field_magnetized}.a. Note that flipping the radial electric field from positive to negative changes both the rotation direction and the fastest and slowest ion population. Heavy (resp. light) ions are thus still pushed outward (resp. inward), so that separation can in principle be achieved with both polarities~\cite{Angerth1962,Anderson1958}.

The first mass separation devices designed to harness these effects were much similar to \emph{homopolar} devices~\cite{Anderson1958}. This configuration consists of a cylindrical anode positioned on the axis of a hollow cylindrical cathode producing a radial electric field. The cathode is surrounded by a set of magnetic coils producing a uniform axial magnetic field in the inter-electrode gap. 
Operation in these devices is typically pulsed, with neutral gas previously fed into the chamber typically ionized by discharging a capacitor bank, but stationary discharges have also been used~\cite{Wijnakker1979}. The discharge current flowing across the magnetic field drives rotation through the Lorentz force $\bm{j}\times\bm{B}$, with $\bm{j}$ the current density~\cite{Lehnert1971}. Experiments confirmed plasma rotation but also revealed different modes of operation depending on the conditions including magnetic field strength $B$ and neutral pressure $p$~\cite{Barber1972}. Furthermore, while these concepts demonstrated separation~\cite{James1976}, the separation factor of partially ionized centrifuges was later shown to be limited by viscous heating~\cite{Wijnakker1980}. The benefits of a higher rotation velocity can then be negated by a higher plasma temperature.

A closely related concept, developed to address the limits of partially ionized centrifuges, is the vacuum arc centrifuge (VAC)~\cite{Krishnan1981}. VACs differ from partially ionized centrifuges in that the chamber is not filled with neutral gas. Instead, the gas discharge is replaced by an arc discharge formed between two electrodes. This allows for \emph{fully} ionized plasmas, which in turn remediates to another limitation of partially ionized centrifuges, namely the critical ionization velocity phenomena originally postulated by Alfv{\'en}~\cite{Alfven1960}. Experiments confirmed the separation capabilities of VACs~\cite{Prasad1987,Hirshfield1989}, but separation factors remain modest even for mass differences of tens of atomic mass units~\cite{Krishnan1983,Poluektov1998}. VACs have also been shown to lead to instabilities~\cite{Hole2002}.

Although rotation is a prerequisite to separation in centrifuges, the condition $E/(Br)\ll\Omega_i$ with $\Omega_i$ the ion cyclotron frequency is typically obtained~\cite{Prasad1987b}, so that centrifugal corrections to the crossed-field rotation frequency can be neglected to lowest order. In addition, ion diamagnetic drift can generally be neglected in front of the $\bm{E}\times\bm{B}$ drift~\cite{Prasad1987b,Bittencourt1987}. Finally, self-generated magnetic fields in partially ionized and vacuum arc centrifuges are generally assumed to be negligible compared to the externally applied magnetic field. Therefore, rotation velocity is primarily controlled by the self-consistent electric field with
\begin{equation}
\Omega\sim \frac{1}{Br}\frac{d\phi}{dr},
\end{equation}
with $\phi$ the electric potential~\cite{Prasad1987b}. Since separation in plasma centrifuges is directly related to the rotation velocity~\cite{Prasad1987a}, the ability to control the electric field and, as a result, the rotation profile, is highly desirable. 

The need for electric field and rotation profile control in magnetized plasma has grown even stronger in the last decade as the emergence of new separation needs~\cite{Gueroult2018} led to the development of new filter concepts. Indeed, the greater mass difference existing between elements to be separated opens new avenues to leverage mass dependent particle dynamics in crossed-field configuration, and configurations which would be inefficient for isotope separation may hold promise for emerging applications~\cite{Fetterman2011b}.


One example is the Archimedes filter~\cite{Freeman2003} based on the DC band gap ion mass filter proposed by Ohkawa and Miller~\cite{Ohkawa2002}.  In this concept, mass separation depends on imposing a suitable DC concave parabolic plasma potential radial profile $\phi_b = \phi_0 (a^2-r^2)$ across a uniform axial magnetic field, with $a$ the axisymmetric plasma column radius. Moving to the frame rotating at the angular frequency $-\Omega_i/2$, the magnetic field cancels. Whether or not an ion is radially confined is then determined by the effective potential $\phi^{\star} = \phi_b+\phi_i$, with
\begin{equation}
\phi_i = \frac{\Omega_iB_0 r^2}{8}
\end{equation}
the contribution of centrifugal and Coriolis forces on a rotating ion~\cite{Gueroult2018}. By imposing a parabolic radial profile, $\phi_b$ has the same radial dependence as $\phi_i$. The voltage drop $\phi_0$ across the plasma column then dictates the sign of $d ^2\phi^{\star}/d r^2$. In particular, it can be chosen so that $\phi^{\star}$ is convex for light ions but concave for heavy ions. If so, light ions are radially confined but heavy ions are not. Light ions could then be collected axially along field lines while heavy ions are collected radially, as shown in Fig.~\ref{Fig:Crossed_field_magnetized}.b. Although conceptually simple, the nature of the separation mechanism in this concept makes high rotation velocity ($\Omega\sim\Omega_i/4$) mandatory, which translates into the need for larger potential gradients. This may in turn bring additional challenges compared to plasma centrifuges, such as the possible onset of instabilities~\cite{Chen1966,Gueroult2017b}. The need for high rotation velocity in a partially ionized plasma may also be an issue due to the critical ionization velocity phenomena~\cite{Brenning1992,LAI2001}. Yet another possible challenge in this concept is that separation relies on single-orbit dynamics. Collisionless operation $\Omega_i/\nu_{ii}\gg 1$, with $\nu_{ii}$ the ion-ion collision frequency, is hence required. This constraint sets in turn an upper density limit for a given magnetic field. Nonetheless, indirect experimental evidence of this differential separation effect was obtained by Shinohara \emph{et al.} in colisionless regime at very low pressure~\cite{Shinohara2007}. Furthermore, an evolution of this concept known as the Double Well Mass Filter, illustrated in Fig.~\ref{Fig:Crossed_field_magnetized}.c, allows for collisional operation and hence possibly for the high-density plasmas required for high-throughput processing~\cite{Gueroult2014}. However, this comes at the expense of the need for a more complex ($4^{\textrm{th}}$ or higher order) electric potential radial profile and shear. Note that while these concepts are considered here under the magnetized ion group, the high rotation velocity makes ion gyro-radii comparable to or even greater than the device radius in the case of heavy ions in the Archimedes filter.

Another separation scheme is the Magnetic Centrifugal Mass Filter (MCMF) proposed by Fetterman and Fisch~\cite{Fetterman2011} drawing upon an asymmetrical centrifugal trap designed for aneutronic fusion~\cite{Volosov1997}. Separation in this device is based on a complex asymmetrical magnetic field topology designed to produce two distinct confinement plugs. A magnetic mirror positioned at larger radius in a centrifugal trap leaks preferentially heavy ions~\cite{Volosov2006,Gueroult2018}, while a centrifugal barrier positioned at a smaller radius preferentially leaks light ions~\cite{Lehnert1974,Bekhtenev1980}. In principle, a heavy ion rich stream can then be collected at one end while a light ion rich stream is collected at the other end. Yet, a suitable electric potential profile needs to be imposed across the magnetic field to produced the plasma rotation required for this device to operate successfully. In addition, ion-ion and ion-neutral collisions constrain the practical range of operation of this concept~\cite{Gueroult2012a,Ochs2017}. While this concept may require less finesse in electric potential profile, in particular compared to Ohkawa's filter and its higher order variation discussed in the previous paragraph, it introduces extra complexity by demanding potential distribution control across magnetic field lines with varying inclination with respect to the rotation axis.

\subsection{Crossed-field in non-magnetized ion regime}
\label{Sec:SecIIb}

Another interesting plasma regime is the one where electrons are magnetized but ions are not. In this regime, electron dynamics consists primarily in the $\bm{E}\times\bm{B}$ drift motion, whereas ion dynamics is primarily dictated by the electric field. This mass and charge dependent dynamics offers many opportunities for applications. This is exemplified by the variety of crossed-field plasma devices operating in this regime, which ranges from magnetrons for sputtering applications to hall-effect thrusters for space propulsion~\cite{Abolmasov2012,Boeuf2017}. In many of these devices, crossed-fields are used to achieve different goals with electron and ion dynamics. For instance, electrons confinement in a closed crossed-field drift geometry ensures efficient ionization while enabling the strong electric field required for ion acceleration in hall-thrusters. But crossed-field configurations in non-magnetized ion regime can also be used to harness the mass dependent ion dynamics.

\begin{figure*}[htbp]
\begin{center}
\includegraphics[]{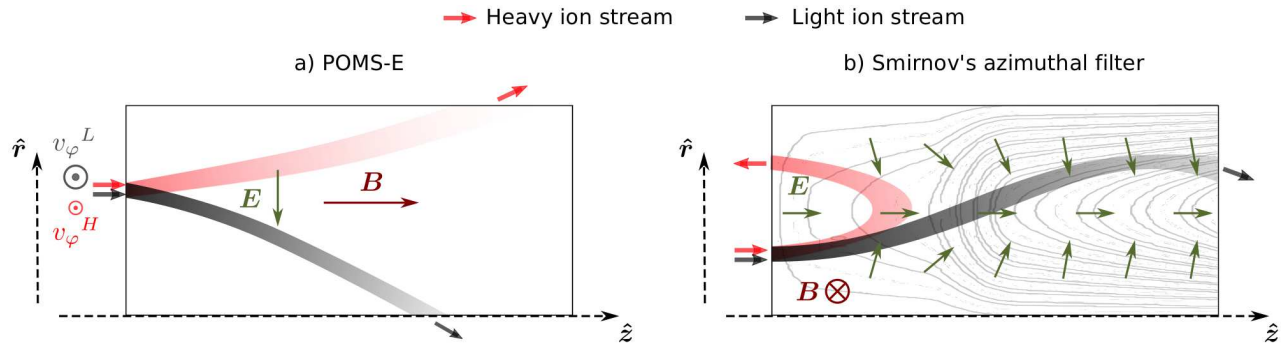}
\caption{Sketch of the separation process in (a) the POMS-E~\cite{Morozov2005,Bardakov2014} and (b) Smironov's azimuthal crossed-field filter~\cite{Smirnov2013}. While ions are non-magnetized in both devices,  ions are affected by the magnetic field in Smirnov's filter while they are not in the POMS-E. POMS-E takes advantage of the balance between centrifugal forces associated with an initial azimuthal velocity (${v_{\varphi}}^H<{v_{\varphi}}^L$) produced upstream in the azimuthator (not shown here) and a confining radial electric field ($E_r<0$).  Smirnov's filter relies on a 2d electric potential well (iso-potential contours in light grey) tailored to accelerate heavy ions axially after light ions have been turned around by the azimuthal magnetic field $\bm{B} = \mu_0I/(2\pi r)\bm{\hat{\varphi}}$, with $I<0$ the on-axis axial current. Thick red and grey arrows represent heavy and light ion flows, respectively.}
\label{Fig:Crossed_field_unmagnetized}
\end{center}
\end{figure*}

One example is the separation region of the Plasma Optical Mass Separator with Electrostatic focusing (POMS-E) originally proposed by Morozov and Savel'ev~\cite{Morozov2005} and further studied and developed by Bardakov and co-workers at Irkutsk State Technical University~\cite{Bardakov2010a,Bardakov2014}. In this concept, an annular plasma beam first passes through a region with strong radial magnetic field known as the ``\emph{azimuthator}''. From conservation of the canonical azimuthal momentum $p_{\varphi} = m v_{\varphi} + q A_{\varphi}$, with $q$ the electric charge, ions acquire an azimuthal momentum $mv_{\varphi}$ upon crossing this region. For a given charge, the azimuthal momentum at the exit of this region $p_{\varphi_a} = mv_{\varphi_a}$ is equal for all ions, so that the azimuthal velocity is inversely proportional to the ion mass. Counter-intuitively, the centrifugal force exerted on ions is hence inversely proportional to the ion mass. Ions then enter the crossed-field separation region which consists, as depicted in Fig.~\ref{Fig:Crossed_field_unmagnetized}.a, in a radial electric field and an axial magnetic field weak enough for ions to be non-magnetized. The ion dynamics in this region can be described by the effective potential
\begin{equation}
\phi^{\ast} = \phi_b - \frac{{p_{\varphi_a}}^2}{m}\ln\left(\frac{r}{r_a}\right)
\end{equation}
with $\phi_b$ the plasma potential and $r_a$ the azimuthator radius. If a negative radial electric field 
\begin{equation}
E_r = -\frac{d\phi_b}{dr} =  -E_0 \frac{r}{r_a}
\end{equation} 
can be imposed in the plasma, with $E_0>0$ the radial electric field at the azimuthator radius, the effective potential can then be rewritten as 
\begin{equation}
\phi^{\ast} = \frac{{p_{\varphi_a}}^2}{e}\left[\frac{1}{m^{\ast}}-\frac{1}{m}\right]\ln\left(\frac{r}{r_a}\right),
\end{equation}
where we have introduced the mass
\begin{equation}
m^{\ast} = \frac{{p_{\varphi_a}}^2}{eE_0r_a}.
\end{equation}

For $m\leq m^{\ast}$, $d \phi^{\ast}/dr\leq 0$, so that lighter ions are pulled outward. Inversely, for $m\geq m^{\ast}$, $d \phi^{\ast}/dr\geq 0$, so that heavier ions are pulled inward. Light ions can then be collected on an outer radial limiter while light ions can be collected on an inner radial limiter, as illustrated in Fig.~\ref{Fig:Crossed_field_unmagnetized}.a. Additionally, an intermediate range of masses ($m\sim m^{\ast}$) can be collected axially by choosing the device length appropriately~\cite{Bardakov2010}. Experiments on this device using a mixture of three gases revealed that the flow becomes positively charged when going through the azimuthator, which in turn affects the ion dynamics and separation~\cite{Bardakov2015}. This observation was recently explained through analytical models for electron and ion flows in the azimuthator~\cite{Bardakov2018}. However, the separation region, and in particular how the required potential can be appropriately imposed across the magnetic field in the presence of plasma, has not yet been studied experimentally. In addition, the extent to which collisions will affect this single-particle separation scheme and how this might constrain the practical operating parameters are unknown to date.

Another plasma mass separation concept based on crossed-fields with non-magnetized ions is that proposed by Smirnov and coworkers at the Joint Institute for High Temperature of the Russian Academy of Sciences and at the Moscow Institute of Technology~\cite{Smirnov2013}. In contrast with other concepts where the magnetic field is mostly axial and typically produced by external coils, the magnetic field is here non-uniform and along the azimuthal direction, $\bm{B} = \mu_0I/(2\pi r)\bm{\hat{\varphi}}$, and is produced by an on-axis axial current $I$. Here $\mu_0$ is the vacuum permittivity and $r$ the radial coordinate. In addition, while ions are non-magnetized, the magnetic field in this concept is strong enough to affect ion dynamics, with $r_c\sim L$. Analytical calculations have shown that mass separation may be produced in different ways in this device, such as axial or radial particle injection and linear or parabolic radial electric potential profile~\cite{Samokhin2016}. For the axial injection scheme illustrated in Fig.~\ref{Fig:Crossed_field_unmagnetized}.b, separation relies on light ions being turned around by the magnetic field while the larger Larmor radius heavy ions reach a potential well which pushes them axially before they can be turned around by the magnetic field. The iso-potential contours of this complex potential topology are traced in light grey in Fig.~\ref{Fig:Crossed_field_unmagnetized}.b. Light ions could then be collected at the same axial position as where they are injected, while heavy ions could be collected further down the beam path. However, control over the imposed potential well depth and position is essential for this device to operate as designed. In addition, similarly to Ohkawa's and the POMS-E filter, separation in this concept relies on single-orbit dynamics, and the effect of collisions typically associated with high density operation on separation capabilities remains to be examined.

\section{Driving crossed-field flows}
\label{Sec:SecIII}
 
As illustrated in Sec.~\ref{Sec:SecII}, mass separation in many concepts is conditioned upon the ability to impose and tailor an electric field perpendicularly to the magnetic field within a plasma. In addition, control over this electric field must be effective for high plasma density. Indeed, while detailed plasma parameters are likely to vary from on concept to another, high-throughput operation requires high density plasmas. Quantitatively, taking a typical device cross sectional area of $1$~m$^2$, and assuming a thermal ion flux $n v_{th,i}/4$ and a typical atomic mass of $60$~amu, one calculates that a plasma density of at least $2~10^{12}$~cm$^{-3}$ is needed to achieve a throughput of $10^4$ kg.yr$^{-1}$ for an ion temperature $T_i\sim10$~eV. It is also worth pointing out here that collisions may limit further the achievable throughput~\cite{Ochs2017}. Consequently, high-throughput plasma mass separation in crossed-field concepts hinges on the demonstration of a means to create and tailor an electric within a plasma for densities of up to $10^{13}$~cm$^{-3}$. In this section, we shed light onto the challenges this represents by reviewing past results. Note though that this is by no means meant to be a comprehensive review of the extensive literature on this topic, but rather an illustration of the complexity of this task in the context of plasma separation.    

\subsection{Foundations for driving crossed-field flows via electrode biasing}

\begin{figure}
\begin{center}
\subfigure[~Limiter]{\includegraphics[]{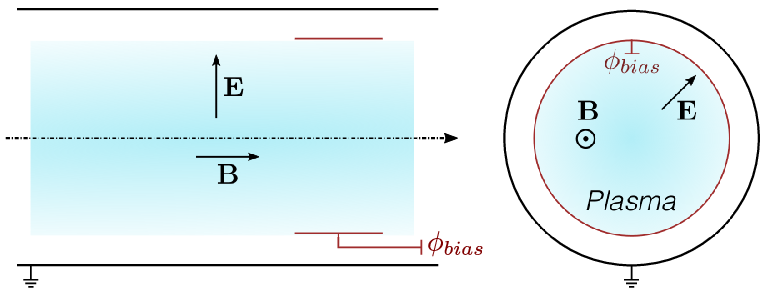}\label{Fig:Biasing_scheme_b}}\\\hspace{0.3cm}\subfigure[~Central rod]{\includegraphics[]{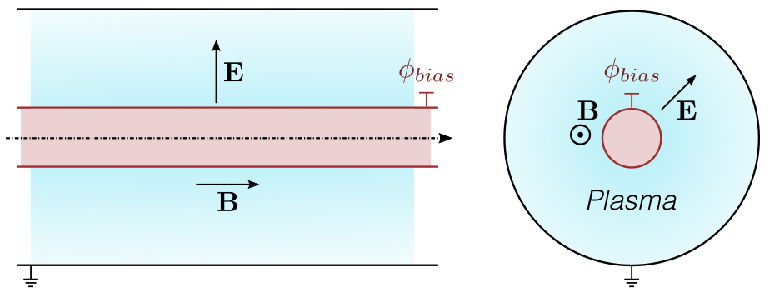}\label{Fig:Biasing_scheme_a}}\\\hspace{-0.1cm}\subfigure[~Ring-electrodes]{\includegraphics[]{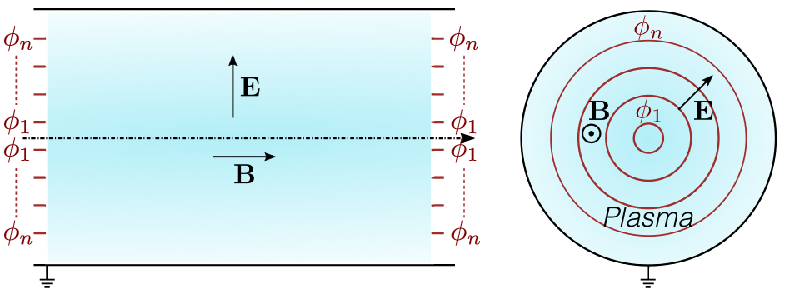}\label{Fig:Biasing_scheme_c}}
\caption{Possible biasing schemes to produce a cross-field configuration: \subref{Fig:Biasing_scheme_b} limiter, \subref{Fig:Biasing_scheme_a} biased central rod and \subref{Fig:Biasing_scheme_c} ring-electrodes. For each scheme, left figure is side view, right figure is end-on view. Black lines denote the cylindrical grounded vacuum vessel, brown lines represent the biased electrodes. }
\vspace{-0.5cm}
\label{Fig:Biasing_schemes}
\end{center}
\end{figure}

In mass filtering concepts, the required electric field is typically assumed to be produced by means of one or multiple biased electrodes within or at the edge of the plasma. However, it is clear that the use of any less than three electrodes will not provide control over the electric field. To illustrate this result, consider for simplicity the linear cylindrical geometry with $\bm{B} = B_0\bm{\hat{z}}$. Positioning a single electrode at the edge of the plasma as illustrated in Fig.~\ref{Fig:Biasing_scheme_b} (often refer to as a \emph{limiter}) can set the local electric potential. It has also been shown to effectively affect the local electric field and rotation in mirror machines~\cite{Bagryansky2003}, linear experiments~\cite{Schaffner2012} and toroidal devices~\cite{Weynants1993}. However, the electric field in the plasma is set through the plasma self-reorganization stemming from the imposed boundary condition. A limiter thus does not provide direct control over the electric field within the plasma column. Similarly, biasing a central rod with respect to an outer electrode, as done in the homopolar configuration~\cite{Anderson1958} depicted in Fig.~\ref{Fig:Biasing_scheme_a}, sets the potential drop across the plasma, but the electric field is still governed by the plasma response. Thus, even if the intricate plasma response were to be entirely predictable, these two schemes could at best provide indirect control over the perpendicular electric field in the plasma.

Another approach is to use biased electrodes along magnetic field lines, such as the ring-electrodes proposed by Lehnert~\cite{Lehnert1970} depicted in Fig.~\ref{Fig:Biasing_scheme_c}. The idea behind this concept is that if conductivity along magnetic field lines can be assumed to be much greater than conductivity perpendicular to field lines, then magnetic field lines are in first approximation iso-potential. The potential of a given field line is then controlled through the potential applied on the electrode intercepting this same field line. This phenomenon is referred to as \emph{magnetic line-tying}~\cite{Lehnert1973}. However, Lehnert pointed out that magnetic line-tying is only effective for a choice of magnetic field and plasma and neutral densities ensuring that collisions with neutrals and viscosity do not alter the velocity field along field lines~\cite{Lehnert1973}. Furthermore, the inter-electrode gap must be smaller than the ion Larmor radius~\cite{Lehnert1974}. These contraints were later underlined by Bekhtenev and co-workers~\cite{Bekhtenev1978,Bekhtenev1980} through the need for high conductivity
between the plasma and electrodes. These conceptual limits are especially relevant for separation applications since high-throughput processing not only requires, as we have seen, high plasma density, but also leads presumably to a non negligible neutral fraction. Indeed, the input feed to be separated is thought to be fed into the system either as dust, droplets or in solid form. It is therefore anticipated that neutrals will be present in the system as a result of partial ionization of the feed. Furthermore, and depending on the feed composition, neutrals are expected to be formed through recycling at the walls.

A fundamental result when considering conductivity in cylindrical geometry is that transport properties are intrinsically modified by rotation~\cite{Helander2005,Rozhansky2008}. Indeed, inertia leads to differences between electron and ion cross-field drift velocities, which in turn makes perpendicular conductivity finite and non-linear in a fully-ionized rotating plasma. Theoretically, Rax \emph{et al.} recently demonstrated that a complex interplay between Coriolis, centrifugal and collisional drag forces makes perpendicular current scale as the third power of the electric field~\cite{Rax2018a}. This non-linear effect is yet another contribution, along with collisions with neutrals and instability and turbulence~\cite{Horton1999}, to the complex picture of transport in crossed-field configurations. Progress towards plasma separation in crossed-field configurations, like very many other plasma applications, thus revolves around advancing our understanding of this complex picture. Pending theoretical advances, insights into the practicality of these concepts can be obtained from experiments.

\subsection{End-electrodes biasing experimental results}

Even if focusing only on experiments in linear geometry, literature on electric potential control using multiple end-electrodes along field lines is quite vast. As illustrated in Tab.~\ref{Ref:Tab1}, experiments can be roughly divided into two groups depending on the scope of the study. On the one hand, there are experiments at low plasma density, low temperature and limited biases conducted to study basic plasma phenomena such as instabilities~\cite{Jassby1972,Komori1988} and space physics~\cite{Amatucci1996}. On the other hand, high density, high temperature strongly biased experiments, conducted for the most part in mirror geometries for magnetic confinement fusion studies~\cite{Mase1991,Abdrashitov1991,Severn1992}. Interestingly, plasma parameters envisioned for separation applications have plasma densities and magnetic field similar to those of the second group but temperatures and neutral fractions more similar to those of the first group. The plasma parameters of the selected experiments discussed below are compared to values targeted for high-throughput separation in Tab.~\ref{Ref:Tab2}.


\begin{table}
\begin{center}
\caption{Typical conditions in past end-electrodes biasing experiments in linear devices. }
\begin{tabular}{c  c  c }
\hline
\hline
\multirow{2}{*}{~Parameter~} & $\quad$Basic plasma$\quad$ & Fusion\\
 & phenomena & experiments  \vspace{0.1cm}\\
\hline
Plasma density $n$ [cm$^{-3}$] & $10^{7}-10^{11}$ & $>10^{12}$ \\
Electron temperature $T_e$ [eV] & $<10$ &  $>30$ \\
Ion temperature $T_i$ [eV] & $10^{-3}-1$ &  $>100$ \\
Neutral density ratio $n_0/n$ & $> 1$ & $<1$ \\
Magnetic field $B$ [kG] & $10^{-3}-1$& $1-10$ \\
Applied bias [V] & $1-100$ & $100-1000$ \\
\hline
\hline
\end{tabular}
\label{Ref:Tab1}
\end{center}
\end{table}

\begin{table*}
\begin{center}
\caption{Typical plasma parameters in selected past end-electrodes biased experiments. Quantities are normalized to the values targeted for high-throughput separation, $n^{\diamond}= 10^{13}$~cm$^{-3}$, $T_e^{\diamond}= 5$~eV, $T_i^{\diamond}= 50$~eV, $E^{\diamond}= 20$~V/cm. }
\begin{tabular}{c  c  c  c  c}
\hline
\hline
 \multirow{3}{*}{~Experiment~} & \multirow{2}{*}{~Plasma density~} & Electron & Ion & \multirow{2}{*}{~Electric field~}\\
 & & temperature & temperature &  \\
 & $[n^{\diamond}]$ & $[T_e^{\diamond}]$ & $[T_i^{\diamond}]$ &  $[E^{\diamond}]$\\
 \hline
Q machine, West Virginia Univ. ~\cite{Caroll1994} &  $10^{-4}$ & $0.01$ & $10^{-3}$ & $0.1$\\ 
Large diameter helicon, Kyushu Univ.~\cite{Shinohara2007} &  $10^{-3}$ & $1$ & $10^{-3}$ & $1$\\
QT-Upgrade machine, Tohoku Univ.~\cite{Tsushima1986,Tsushima1991} &  $0.02$ & $1$ & $0.02$ & $0.1$\\
Gamma 10 mirror, Univ. Tsukuba~\cite{Mase1991} &  $0.1$ & $15$ & $50$ & $10$\\ 
LAPD afterglow, Univ. California Los Angeles~\cite{Koepke2008} &  $0.1$ & $0.1$ & $0.01$ & $0.3$\\ 
Phaedrus tandem mirror, Univ. Wisconsin~\cite{Severn1992} &  $0.2$ & $3$ & $0.6$ & $1$\\ 
Helcat, Univ. New Mexico~\cite{Gilmore2009} &  $1$ & $1$ & $0.02$ & $0.1$\\ 
PMFX, Princeton Plasma Physics Lab.~\cite{Gueroult2016a} &  $1$ & $1$ & $0.02$ & $0.1$\\
C-2 device, Tri Alpha Energy~\cite{Tuszewski2012}&  $4$ & $100$ & $50$ & $0.5$\\
\hline
\hline
\end{tabular}
\label{Ref:Tab2}
\end{center}
\end{table*}

Among low density experiments, the most successful is arguably the experiment by Tsushima and co-workers using three concentric ring-electrodes in an argon electron cyclotron resonance plasma in mirror geometry~\cite{Tsushima1986,Tsushima1991}. The radial electric field near the mirror midplane was successfully varied from $-2$ to $+2$~V/cm at $n\sim10^{11}$~cm$^{-3}$ by applying biases of no more than $\pm20~$V on electrodes located in the end cells $70$~cm away down and up the magnetic field lines. From the operating pressure, the neutral density $n_0$ in this experiment is about $10$ times the plasma density. Another relevant biasing experiment is the one by Shinohara and Horii~\cite{Shinohara2007} in a low-pressure radio-frequency (RF) plasma ($n\sim10^{10}$~cm$^{-3}$). In this experiment, only a step potential profile was applied to ring-electrodes and no attempt was made to examine negative radial electric fields. Yet, it showed that applying a single positive bias on a set of ring electrodes leads to a uniform increase of the floating potential within the radius of the innermost biased electrode. Floating potential about half the applied bias were measured ($80$~V for a $200$~V bias). Consistent with the change in floating potential radial profile, a localised positive radial electric field is formed. More importantly, Shinohara and Horii~\cite{Shinohara2007} showed indirect evidence of mass differential confinement which is consistent with Ohkawa's collisionless separation scheme~\cite{Ohkawa2002} introduced in Sec.~\ref{Sec:SecIIa} and illustrated in Fig.~\ref{Fig:Crossed_field_magnetized}.b. Consistent with these results, localised positive radial electric fields have been been reported when positively biasing the inner electrode of a two electrodes set in a Q-machine~\cite{Caroll1994}. In a similar configuration, negative electric fields were produced with localised negative bias using emissive electrodes~\cite{Jassby1972}. Recently, Liziakin \emph{et al.} showed that end-electrodes effectiveness to affect the potential increased with the electrodes' area, both in reflex~\cite{Liziakin2016} and RF~\cite{Liziakin2017} discharges. 

At high plasma density, Mase \emph{et al.} demonstrated that the electric field in the central cell of the $27$~m long GAMMA 10 mirror experiment could be varied both in strength and polarity (from $-20$~V/cm to $175$~V/cm) by applying kV biases on end electrodes positioned in the end-cells of the machine~\cite{Mase1991}. Similarly to Shinohara's results at lower density, Severn and Hershkowitz showed that a positive bias on a ring electrode leads to a uniform increase of the floating potential within the radius of this electrode, which makes it possible to double the electric field at the edge~\cite{Severn1992}. Still at high density ($\sim10^{13}$~cm$^{-3}$), Tuszewski \emph{et al.} recently measured that the potential imposed on electrodes at one end of the $18$~m long C-2 device is recovered at the other end, indicating that potential is transmitted efficiently along the magnetic field lines going through the edge layer of this field reverse configuration (FRC) plasma~\cite{Tuszewski2012}. It is worth noting here that the interpretation of results obtained in mirror machines and FRCs is made even more complex by the fact that plasma parameters (density, magnetic field and temperature) vary strongly along a given field line in between end-electrodes.  

The end-electrode biasing data with plasma parameters closest to those required for plasma separation are arguably those obtained by Gilmore \emph{et al.} in the HelCat helicon plasma ($n\sim 10^{13}$) with a single set of ring-electrodes~\cite{Gilmore2014}. However, owing to the different scope of this experiment (turbulence suppression), only data on the effect of a global biasing of the electrodes set are available. These results indicate little effect in case of negative biasing (as low as $-16$~V)~\cite{Gilmore2009}. On the other hand, positive biasing up to $20$~V leads to a roughly uniform increase of the plasma potential for radii smaller than the largest electrode radius~\cite{Gilmore2009}.  Similarly, imposing a radial voltage drop of up to $150$~V across a single set of ring electrodes in an afterglow discharge of the Large Plasma Device has been showed to lead to a progressive increase of the potential in the core region~\cite{Koepke2008}. These different results for positive biases are qualitatively consistent with results at low density discussed earlier. Experiments at PMFX with individual biasing of three ring-electrodes in similar plasma conditions suggested that the plasma potential gradient may indeed follow locally the potential imposed on the electrodes~\cite{Gueroult2016a}. However, this result ought to be confirmed, if possible with symmetric boundary conditions and a uniform magnetic field.

In summary, there seems to exist corroborating evidence across a wide range of plasma density that biasing positively a ring electrode leads to a uniform increase of the plasma potential within this ring. This can in turn be used to affect the positive radial electric field near the radius of this ring. However, with the exception of the work by Tsushima and co-workers~\cite{Tsushima1986,Tsushima1991} and, to a lesser extent, by Koepke~\emph{et al.}\cite{Koepke2008}, the ability to combine multiple rings to tailor the electric field profile throughout the plasma column has not yet been demonstrated. In addition, the contrasting results obtained with negative biases by Gilmore \emph{et al.}~\cite{Gilmore2009} on the one hand and Mase \emph{et al.}~\cite{Mase1991} on the other hand brings into question the practicality of forming a negative electric field (radial potential well) such as the one needed in the POMS-E concept (see Sec.~\ref{Sec:SecIIb} and Fig.~\ref{Fig:Crossed_field_unmagnetized}.a) using end-electrodes. Results from Jassby~\cite{Jassby1972} suggest that electron emission at the electrodes may be important in this case.

\subsection{Promising alternative crossed-field flow driving mechanism}

As a substitute for end-electrode biasing, Fetterman and Fisch suggested using waves to drive the $\bm{E}\times\bm{B}$ flow~\cite{Fetterman2008}. Here, the radial electric field is obtained as a result of the charge separation induced by the injection of waves with azimuthal phase velocities in the plasma~\cite{Fisch1992}. This field can in turn produce rotation~\cite{Fetterman2009}. This angular momentum transfer between wave and particle is governed by wave-particle resonant conditions~\cite{Fisch1992,Fisch1992a}. For a wave with azimuthal mode number $l$ and an ion at radial position $r$ with energy $\varepsilon$ and guiding center orbital angular momentum $L = m_i\Omega r^2$, the transfer of a quantum of energy $\delta \varepsilon = \hbar \omega$ from the wave to the ion is associated with the transfer of a quantum of angular momentum $\delta L = \hbar l$ from the wave to this same ion.  The classical ratio 
$\delta L/\delta \varepsilon=l/\omega$ points to the use of high $l$ modes at low frequency.

While this control scheme has not been experimentally demonstrated, it offers conceptual advantages over electrode biasing. In particular, the absence of electrodes in contact with the plasma is advantageous for many reasons. Besides possible electrode oxydation or contamination, contact between the electrodes and the plasma also leads to the presence of neutrals as a result of particle recycling. This neutral fraction can in turn favor the the onset of the critical ionization phenomena~\cite{Alfven1960}, which then limits the practical plasma rotation velocity. Retiring the need for electrodes by relying on wave driven cross-field flow would thus circumvent this potential challenge. Another motivation to study this driving scheme is that it holds significant promise beyond plasma separation. Indeed, wave driven rotation has been suggested to confine particles in toroidal geometry without the need for a poloidal magnetic field~\cite{Rax2017,Ochs2017b}.

\section{Summary}
\label{Sec:SecIV}

There exists an increasing number of societal challenges for which conventional separation techniques are either inefficient or suffer from significant downsides. In contrast, it has been shown that high-throughput separation based on mass at the elemental level is uniquely suited to address these needs. Plasma mass separation hence appear to be a promising a solution. Yet, these emerging needs stand out in that they require separating elements with large mass differences at high-throughput.  Such capabilities are typically beyond those offered by plasma separator concepts previously developed for isotope separation. New concepts are therefore called for.


Crossed-field configurations offer many opportunities for mass separation. In regimes where ions are magnetized, crossed-field configurations can in principle be designed to produce plasma rotation through a suitable drift motion. Inertial effects associated with rotation can then be used in various ways to separate elements based on mass. Conversely, in regimes where ions are unmagnetized, the mass dependent ion dynamics of ion beams injected in crossed-field configurations can be used for mass separation. However, a key requirement in any of these concepts is the ability to impose a suitable electric field perpendicular to the magnetic field within the plasma. 

An often suggested solution to impose and control an electric field perpendicular to the magnetic field is to use biased end-electrodes. However, the efficiency of this scheme relies, among other things, on the perpendicular conductivity being negligible compared to parallel conductivity. Yet, a comprehensive picture of the interplay between the various phenomena driving perpendicular conductivity in crossed-field configurations is still missing. Assessing the practicality of end-electrodes biasing for high-throughput plasma separation  therefore remains beyond current predictive capabilities. Nevertheless, past experimental results obtained in different machines and operating conditions suggest that end-electrodes can indeed, under some conditions, effectively create a radially localised positive radial electric field in linear machines. It remains though to demonstrate this capability unequivocally for plasma parameters relevant to high-throughput plasma separation. Progress towards this goal hinges on adavancing our understanding of crossed-field conductivity, both through experiments and theory.

\section*{Acknowledgments} 
The authors would like to thank I E Ochs, E J Kolmes and M Mlodik for constructive discussions.

\section*{References} 

\begin{thebibliography}{107}%
\makeatletter
\providecommand \@ifxundefined [1]{%
 \@ifx{#1\undefined}
}%
\providecommand \@ifnum [1]{%
 \ifnum #1\expandafter \@firstoftwo
 \else \expandafter \@secondoftwo
 \fi
}%
\providecommand \@ifx [1]{%
 \ifx #1\expandafter \@firstoftwo
 \else \expandafter \@secondoftwo
 \fi
}%
\providecommand \natexlab [1]{#1}%
\providecommand \enquote  [1]{``#1''}%
\providecommand \bibnamefont  [1]{#1}%
\providecommand \bibfnamefont [1]{#1}%
\providecommand \citenamefont [1]{#1}%
\providecommand \href@noop [0]{\@secondoftwo}%
\providecommand \href [0]{\begingroup \@sanitize@url \@href}%
\providecommand \@href[1]{\@@startlink{#1}\@@href}%
\providecommand \@@href[1]{\endgroup#1\@@endlink}%
\providecommand \@sanitize@url [0]{\catcode `\\12\catcode `\$12\catcode
  `\&12\catcode `\#12\catcode `\^12\catcode `\_12\catcode `\%12\relax}%
\providecommand \@@startlink[1]{}%
\providecommand \@@endlink[0]{}%
\providecommand \url  [0]{\begingroup\@sanitize@url \@url }%
\providecommand \@url [1]{\endgroup\@href {#1}{\urlprefix }}%
\providecommand \urlprefix  [0]{URL }%
\providecommand \Eprint [0]{\href }%
\providecommand \doibase [0]{http://dx.doi.org/}%
\providecommand \selectlanguage [0]{\@gobble}%
\providecommand \bibinfo  [0]{\@secondoftwo}%
\providecommand \bibfield  [0]{\@secondoftwo}%
\providecommand \translation [1]{[#1]}%
\providecommand \BibitemOpen [0]{}%
\providecommand \bibitemStop [0]{}%
\providecommand \bibitemNoStop [0]{.\EOS\space}%
\providecommand \EOS [0]{\spacefactor3000\relax}%
\providecommand \BibitemShut  [1]{\csname bibitem#1\endcsname}%
\let\auto@bib@innerbib\@empty
\bibitem [{\citenamefont {Cussler}\ and\ \citenamefont
  {Dutta}(2012)}]{Cussler2012}%
  \BibitemOpen
  \bibfield  {author} {\bibinfo {author} {\bibfnamefont {E.~L.}\ \bibnamefont
  {Cussler}}\ and\ \bibinfo {author} {\bibfnamefont {B.~K.}\ \bibnamefont
  {Dutta}},\ }\href {\doibase 10.1002/aic.13779} {\bibfield  {journal}
  {\bibinfo  {journal} {{AIChE} Journal}\ }\textbf {\bibinfo {volume} {58}},\
  \bibinfo {pages} {3825} (\bibinfo {year} {2012})}\BibitemShut {NoStop}%
\bibitem [{\citenamefont {Sholl}\ and\ \citenamefont
  {Lively}(2016)}]{Sholl2016}%
  \BibitemOpen
  \bibfield  {author} {\bibinfo {author} {\bibfnamefont {D.~S.}\ \bibnamefont
  {Sholl}}\ and\ \bibinfo {author} {\bibfnamefont {R.~P.}\ \bibnamefont
  {Lively}},\ }\href {\doibase 10.1038/532435a} {\bibfield  {journal} {\bibinfo
   {journal} {Nature}\ }\textbf {\bibinfo {volume} {532}},\ \bibinfo {pages}
  {435} (\bibinfo {year} {2016})}\BibitemShut {NoStop}%
\bibitem [{\citenamefont {Jou}\ \emph {et~al.}(2014)\citenamefont {Jou},
  \citenamefont {Melnikov}, \citenamefont {Nadtochiy},\ and\ \citenamefont
  {Gracheva}}]{Jou2014}%
  \BibitemOpen
  \bibfield  {author} {\bibinfo {author} {\bibfnamefont {I.~A.}\ \bibnamefont
  {Jou}}, \bibinfo {author} {\bibfnamefont {D.~V.}\ \bibnamefont {Melnikov}},
  \bibinfo {author} {\bibfnamefont {A.}~\bibnamefont {Nadtochiy}}, \ and\
  \bibinfo {author} {\bibfnamefont {M.~E.}\ \bibnamefont {Gracheva}},\ }\href
  {\doibase 10.1088/0957-4484/25/14/145201} {\bibfield  {journal} {\bibinfo
  {journal} {Nanotechnology}\ }\textbf {\bibinfo {volume} {25}},\ \bibinfo
  {pages} {145201} (\bibinfo {year} {2014})}\BibitemShut {NoStop}%
\bibitem [{\citenamefont {Kang}\ and\ \citenamefont {Cannon}(2015)}]{Kang2015}%
  \BibitemOpen
  \bibfield  {author} {\bibinfo {author} {\bibfnamefont {W.}~\bibnamefont
  {Kang}}\ and\ \bibinfo {author} {\bibfnamefont {J.~L.}\ \bibnamefont
  {Cannon}},\ }\href {\doibase 10.1371/journal.pone.0141484} {\bibfield
  {journal} {\bibinfo  {journal} {{PLOS} {ONE}}\ }\textbf {\bibinfo {volume}
  {10}},\ \bibinfo {pages} {e0141484} (\bibinfo {year} {2015})}\BibitemShut
  {NoStop}%
\bibitem [{\citenamefont {Dempster}(1918)}]{Dempster1918}%
  \BibitemOpen
  \bibfield  {author} {\bibinfo {author} {\bibfnamefont {A.~J.}\ \bibnamefont
  {Dempster}},\ }\href {\doibase 10.1103/physrev.11.316} {\bibfield  {journal}
  {\bibinfo  {journal} {Phys. Rev.}\ }\textbf {\bibinfo {volume} {11}},\
  \bibinfo {pages} {316} (\bibinfo {year} {1918})}\BibitemShut {NoStop}%
\bibitem [{\citenamefont {Aston}(1919)}]{Aston1919}%
  \BibitemOpen
  \bibfield  {author} {\bibinfo {author} {\bibfnamefont {F.~W.}\ \bibnamefont
  {Aston}},\ }\href {\doibase 10.1080/14786441208636004} {\bibfield  {journal}
  {\bibinfo  {journal} {Lond. Edinb. Dubl. Phil. Mag.}\ }\textbf {\bibinfo
  {volume} {38}},\ \bibinfo {pages} {707} (\bibinfo {year} {1919})}\BibitemShut
  {NoStop}%
\bibitem [{\citenamefont {Wien}(1898)}]{Wien1898}%
  \BibitemOpen
  \bibfield  {author} {\bibinfo {author} {\bibfnamefont {W.}~\bibnamefont
  {Wien}},\ }\href@noop {} {\bibfield  {journal} {\bibinfo  {journal} {Ann.
  Phys.}\ }\textbf {\bibinfo {volume} {65}},\ \bibinfo {pages} {440} (\bibinfo
  {year} {1898})}\BibitemShut {NoStop}%
\bibitem [{\citenamefont {Thomson}(1913)}]{Thomson1913}%
  \BibitemOpen
  \bibfield  {author} {\bibinfo {author} {\bibfnamefont {J.~J.}\ \bibnamefont
  {Thomson}},\ }\href {\doibase 10.1098/rspa.1913.0057} {\bibfield  {journal}
  {\bibinfo  {journal} {Proc. Royal Soc. Lond. A.}\ }\textbf {\bibinfo {volume}
  {89}},\ \bibinfo {pages} {1} (\bibinfo {year} {1913})}\BibitemShut {NoStop}%
\bibitem [{\citenamefont {Beynon}\ and\ \citenamefont
  {Morgan}(1978)}]{Beynon1978}%
  \BibitemOpen
  \bibfield  {author} {\bibinfo {author} {\bibfnamefont {J.}~\bibnamefont
  {Beynon}}\ and\ \bibinfo {author} {\bibfnamefont {R.}~\bibnamefont
  {Morgan}},\ }\href {\doibase 10.1016/0020-7381(78)80098-9} {\bibfield
  {journal} {\bibinfo  {journal} {Int. J. Mass Spectrom. Ion Phys.}\ }\textbf
  {\bibinfo {volume} {27}},\ \bibinfo {pages} {1} (\bibinfo {year}
  {1978})}\BibitemShut {NoStop}%
\bibitem [{\citenamefont {M{\"u}nzenberg}(2013)}]{Muenzenberg2013}%
  \BibitemOpen
  \bibfield  {author} {\bibinfo {author} {\bibfnamefont {G.}~\bibnamefont
  {M{\"u}nzenberg}},\ }\href {\doibase 10.1016/j.ijms.2013.03.009} {\bibfield
  {journal} {\bibinfo  {journal} {Int. J. Mass Spectrom.}\ }\textbf {\bibinfo
  {volume} {349-350}},\ \bibinfo {pages} {9} (\bibinfo {year}
  {2013})}\BibitemShut {NoStop}%
\bibitem [{\citenamefont {Lawrence}(1958)}]{Lawrence1958}%
  \BibitemOpen
  \bibfield  {author} {\bibinfo {author} {\bibfnamefont {E.~O.}\ \bibnamefont
  {Lawrence}},\ }\href@noop {} {\enquote {\bibinfo {title} {Calutron system},}\
  } (\bibinfo {year} {1958})\BibitemShut {NoStop}%
\bibitem [{\citenamefont {Alexeff}(1978)}]{Alexeff1978}%
  \BibitemOpen
  \bibfield  {author} {\bibinfo {author} {\bibfnamefont {I.}~\bibnamefont
  {Alexeff}},\ }\href {\doibase 10.1109/tps.1978.4317160} {\bibfield  {journal}
  {\bibinfo  {journal} {{IEEE} Trans. Plasma Science}\ }\textbf {\bibinfo
  {volume} {6}},\ \bibinfo {pages} {539} (\bibinfo {year} {1978})}\BibitemShut
  {NoStop}%
\bibitem [{\citenamefont {Smith}, \citenamefont {Parkins},\ and\ \citenamefont
  {Forrester}(1947)}]{Smith1947}%
  \BibitemOpen
  \bibfield  {author} {\bibinfo {author} {\bibfnamefont {L.~P.}\ \bibnamefont
  {Smith}}, \bibinfo {author} {\bibfnamefont {W.~E.}\ \bibnamefont {Parkins}},
  \ and\ \bibinfo {author} {\bibfnamefont {A.~T.}\ \bibnamefont {Forrester}},\
  }\href {\doibase 10.1103/physrev.72.989} {\bibfield  {journal} {\bibinfo
  {journal} {Phys. Rev.}\ }\textbf {\bibinfo {volume} {72}},\ \bibinfo {pages}
  {989} (\bibinfo {year} {1947})}\BibitemShut {NoStop}%
\bibitem [{\citenamefont {Parkins}(2005)}]{Parkins2005}%
  \BibitemOpen
  \bibfield  {author} {\bibinfo {author} {\bibfnamefont {W.~E.}\ \bibnamefont
  {Parkins}},\ }\href {\doibase 10.1063/1.1995747} {\bibfield  {journal}
  {\bibinfo  {journal} {Phys. Today}\ }\textbf {\bibinfo {volume} {58}},\
  \bibinfo {pages} {45} (\bibinfo {year} {2005})}\BibitemShut {NoStop}%
\bibitem [{\citenamefont {Boeschoten}\ and\ \citenamefont
  {Nathrath}(1979)}]{Boeschoten1979}%
  \BibitemOpen
  \bibfield  {author} {\bibinfo {author} {\bibfnamefont {F.}~\bibnamefont
  {Boeschoten}}\ and\ \bibinfo {author} {\bibfnamefont {N.}~\bibnamefont
  {Nathrath}},\ }\enquote {\bibinfo {title} {Plasma separating effects},}\ in\
  \href {\doibase 10.1007/3-540-09385-0_21} {\emph {\bibinfo {booktitle}
  {Uranium Enrichment}}},\ \bibinfo {editor} {edited by\ \bibinfo {editor}
  {\bibfnamefont {S.}~\bibnamefont {Villani}}}\ (\bibinfo  {publisher}
  {Springer Berlin Heidelberg},\ \bibinfo {address} {Berlin, Heidelberg},\
  \bibinfo {year} {1979})\ pp.\ \bibinfo {pages} {291--315}\BibitemShut
  {NoStop}%
\bibitem [{\citenamefont {Louvet}(1989)}]{Louvet1989}%
  \BibitemOpen
  \bibfield  {author} {\bibinfo {author} {\bibfnamefont {P.}~\bibnamefont
  {Louvet}},\ }in\ \href@noop {} {\emph {\bibinfo {booktitle} {Proceedings of
  the 2nd workshop on Separation Phenomena in Liquids and Gases}}}\ (\bibinfo
  {year} {1989})\ pp.\ \bibinfo {pages} {5--104}\BibitemShut {NoStop}%
\bibitem [{\citenamefont {Grossman}\ and\ \citenamefont
  {Shepp}(1991)}]{Grossman1991}%
  \BibitemOpen
  \bibfield  {author} {\bibinfo {author} {\bibfnamefont {M.~W.}\ \bibnamefont
  {Grossman}}\ and\ \bibinfo {author} {\bibfnamefont {T.~A.}\ \bibnamefont
  {Shepp}},\ }\href {\doibase 10.1109/27.125034} {\bibfield  {journal}
  {\bibinfo  {journal} {IEEE Trans. Plasma Sci.}\ }\textbf {\bibinfo {volume}
  {19}},\ \bibinfo {pages} {1114} (\bibinfo {year} {1991})}\BibitemShut
  {NoStop}%
\bibitem [{\citenamefont {Dolgolenko}\ and\ \citenamefont
  {Muromkin}(2017)}]{Dolgolenko2017}%
  \BibitemOpen
  \bibfield  {author} {\bibinfo {author} {\bibfnamefont {D.~A.}\ \bibnamefont
  {Dolgolenko}}\ and\ \bibinfo {author} {\bibfnamefont {Y.~A.}\ \bibnamefont
  {Muromkin}},\ }\href {\doibase 10.3367/UFNe.2016.12.038016} {\bibfield
  {journal} {\bibinfo  {journal} {Phys. Usp.}\ }\textbf {\bibinfo {volume}
  {60}},\ \bibinfo {pages} {994} (\bibinfo {year} {2017})}\BibitemShut
  {NoStop}%
\bibitem [{\citenamefont {Gueroult}, \citenamefont {Hobbs},\ and\ \citenamefont
  {Fisch}(2015)}]{Gueroult2015}%
  \BibitemOpen
  \bibfield  {author} {\bibinfo {author} {\bibfnamefont {R.}~\bibnamefont
  {Gueroult}}, \bibinfo {author} {\bibfnamefont {D.~T.}\ \bibnamefont {Hobbs}},
  \ and\ \bibinfo {author} {\bibfnamefont {N.~J.}\ \bibnamefont {Fisch}},\
  }\href {\doibase 10.1016/j.jhazmat.2015.04.058} {\bibfield  {journal}
  {\bibinfo  {journal} {J. Hazard. Mater.}\ }\textbf {\bibinfo {volume}
  {297}},\ \bibinfo {pages} {153} (\bibinfo {year} {2015})}\BibitemShut
  {NoStop}%
\bibitem [{\citenamefont {Timofeev}(2014)}]{Timofeev2014}%
  \BibitemOpen
  \bibfield  {author} {\bibinfo {author} {\bibfnamefont {A.~V.}\ \bibnamefont
  {Timofeev}},\ }\href {\doibase 10.3367/UFNe.0184.201410g.1101} {\bibfield
  {journal} {\bibinfo  {journal} {Sov. Phys. Usp.}\ }\textbf {\bibinfo {volume}
  {57}},\ \bibinfo {pages} {990} (\bibinfo {year} {2014})}\BibitemShut
  {NoStop}%
\bibitem [{\citenamefont {Gueroult}\ and\ \citenamefont
  {Fisch}(2014)}]{Gueroult2014a}%
  \BibitemOpen
  \bibfield  {author} {\bibinfo {author} {\bibfnamefont {R.}~\bibnamefont
  {Gueroult}}\ and\ \bibinfo {author} {\bibfnamefont {N.~J.}\ \bibnamefont
  {Fisch}},\ }\href {\doibase 10.1088/0963-0252/23/3/035002} {\bibfield
  {journal} {\bibinfo  {journal} {Plasma Sources Sci. Technol.}\ }\textbf
  {\bibinfo {volume} {23}},\ \bibinfo {pages} {035002} (\bibinfo {year}
  {2014})}\BibitemShut {NoStop}%
\bibitem [{\citenamefont {Vorona}\ \emph {et~al.}(2015)\citenamefont {Vorona},
  \citenamefont {Gavrikov}, \citenamefont {Samokhin}, \citenamefont {Smirnov},\
  and\ \citenamefont {Khomyakov}}]{Vorona2015}%
  \BibitemOpen
  \bibfield  {author} {\bibinfo {author} {\bibfnamefont {N.~A.}\ \bibnamefont
  {Vorona}}, \bibinfo {author} {\bibfnamefont {A.~V.}\ \bibnamefont
  {Gavrikov}}, \bibinfo {author} {\bibfnamefont {A.~A.}\ \bibnamefont
  {Samokhin}}, \bibinfo {author} {\bibfnamefont {V.~P.}\ \bibnamefont
  {Smirnov}}, \ and\ \bibinfo {author} {\bibfnamefont {Y.~S.}\ \bibnamefont
  {Khomyakov}},\ }\href {http://dx.doi.org/10.1134/S1063778815140148}
  {\bibfield  {journal} {\bibinfo  {journal} {Phys. At. Nucl.}\ }\textbf
  {\bibinfo {volume} {78}},\ \bibinfo {pages} {1624} (\bibinfo {year}
  {2015})}\BibitemShut {NoStop}%
\bibitem [{\citenamefont {Yuferov}\ \emph {et~al.}(2017)\citenamefont
  {Yuferov}, \citenamefont {Shariy}, \citenamefont {Tkachova}, \citenamefont
  {Katrechko}, \citenamefont {Svichkar}, \citenamefont {Ilichova},
  \citenamefont {Shvets},\ and\ \citenamefont {Mufel}}]{Yuferov2017}%
  \BibitemOpen
  \bibfield  {author} {\bibinfo {author} {\bibfnamefont {V.~B.}\ \bibnamefont
  {Yuferov}}, \bibinfo {author} {\bibfnamefont {S.~V.}\ \bibnamefont {Shariy}},
  \bibinfo {author} {\bibfnamefont {T.~I.}\ \bibnamefont {Tkachova}}, \bibinfo
  {author} {\bibfnamefont {V.~V.}\ \bibnamefont {Katrechko}}, \bibinfo {author}
  {\bibfnamefont {A.~S.}\ \bibnamefont {Svichkar}}, \bibinfo {author}
  {\bibfnamefont {V.~O.}\ \bibnamefont {Ilichova}}, \bibinfo {author}
  {\bibfnamefont {M.~O.}\ \bibnamefont {Shvets}}, \ and\ \bibinfo {author}
  {\bibfnamefont {E.~V.}\ \bibnamefont {Mufel}},\ }\href@noop {} {\bibfield
  {journal} {\bibinfo  {journal} {Problems Atomic Sci. Tech. Ser.: Plasma
  physics}\ }\textbf {\bibinfo {volume} {107}},\ \bibinfo {pages} {223}
  (\bibinfo {year} {2017})}\BibitemShut {NoStop}%
\bibitem [{\citenamefont {Gueroult}, \citenamefont {Rax},\ and\ \citenamefont
  {Fisch}(2018)}]{Gueroult2018a}%
  \BibitemOpen
  \bibfield  {author} {\bibinfo {author} {\bibfnamefont {R.}~\bibnamefont
  {Gueroult}}, \bibinfo {author} {\bibfnamefont {J.~M.}\ \bibnamefont {Rax}}, \
  and\ \bibinfo {author} {\bibfnamefont {N.~J.}\ \bibnamefont {Fisch}},\ }\href
  {\doibase 10.1016/j.jclepro.2018.02.066} {\bibfield  {journal} {\bibinfo
  {journal} {J. Clean. Prod.}\ }\textbf {\bibinfo {volume} {182}},\ \bibinfo
  {pages} {1060} (\bibinfo {year} {2018})}\BibitemShut {NoStop}%
\bibitem [{\citenamefont {Gueroult}\ \emph {et~al.}(2018)\citenamefont
  {Gueroult}, \citenamefont {Rax}, \citenamefont {Zweben},\ and\ \citenamefont
  {Fisch}}]{Gueroult2018}%
  \BibitemOpen
  \bibfield  {author} {\bibinfo {author} {\bibfnamefont {R.}~\bibnamefont
  {Gueroult}}, \bibinfo {author} {\bibfnamefont {J.-M.}\ \bibnamefont {Rax}},
  \bibinfo {author} {\bibfnamefont {S.}~\bibnamefont {Zweben}}, \ and\ \bibinfo
  {author} {\bibfnamefont {N.~J.}\ \bibnamefont {Fisch}},\ }\href {\doibase
  10.1088/1361-6587/aa8be5} {\bibfield  {journal} {\bibinfo  {journal} {Plasma
  Phys. Control. Fusion}\ }\textbf {\bibinfo {volume} {60}},\ \bibinfo {pages}
  {014018} (\bibinfo {year} {2018})}\BibitemShut {NoStop}%
\bibitem [{\citenamefont {Zweben}, \citenamefont {Gueroult},\ and\
  \citenamefont {Fisch}(2018)}]{Zweben2018}%
  \BibitemOpen
  \bibfield  {author} {\bibinfo {author} {\bibfnamefont {S.~J.}\ \bibnamefont
  {Zweben}}, \bibinfo {author} {\bibfnamefont {R.}~\bibnamefont {Gueroult}}, \
  and\ \bibinfo {author} {\bibfnamefont {N.~J.}\ \bibnamefont {Fisch}},\ }\href
  {\doibase 10.1063/1.5042845} {\bibfield  {journal} {\bibinfo  {journal}
  {Phys. Plasmas}\ }\textbf {\bibinfo {volume} {25}},\ \bibinfo {pages}
  {090901} (\bibinfo {year} {2018})}\BibitemShut {NoStop}%
\bibitem [{\citenamefont {Dawson}\ \emph {et~al.}(1976)\citenamefont {Dawson},
  \citenamefont {Kim}, \citenamefont {Arnush}, \citenamefont {Fried},
  \citenamefont {Gould}, \citenamefont {Heflinger}, \citenamefont {Kennel},
  \citenamefont {Romesser}, \citenamefont {Stenzel}, \citenamefont {Wong},\
  and\ \citenamefont {Wuerker}}]{Dawson1976}%
  \BibitemOpen
  \bibfield  {author} {\bibinfo {author} {\bibfnamefont {J.~M.}\ \bibnamefont
  {Dawson}}, \bibinfo {author} {\bibfnamefont {H.~C.}\ \bibnamefont {Kim}},
  \bibinfo {author} {\bibfnamefont {D.}~\bibnamefont {Arnush}}, \bibinfo
  {author} {\bibfnamefont {B.~D.}\ \bibnamefont {Fried}}, \bibinfo {author}
  {\bibfnamefont {R.~W.}\ \bibnamefont {Gould}}, \bibinfo {author}
  {\bibfnamefont {L.~O.}\ \bibnamefont {Heflinger}}, \bibinfo {author}
  {\bibfnamefont {C.~F.}\ \bibnamefont {Kennel}}, \bibinfo {author}
  {\bibfnamefont {T.~E.}\ \bibnamefont {Romesser}}, \bibinfo {author}
  {\bibfnamefont {R.~L.}\ \bibnamefont {Stenzel}}, \bibinfo {author}
  {\bibfnamefont {A.~Y.}\ \bibnamefont {Wong}}, \ and\ \bibinfo {author}
  {\bibfnamefont {R.~F.}\ \bibnamefont {Wuerker}},\ }\href {\doibase
  10.1103/physrevlett.37.1547} {\bibfield  {journal} {\bibinfo  {journal}
  {Phys. Rev. Lett.}\ }\textbf {\bibinfo {volume} {37}},\ \bibinfo {pages}
  {1547} (\bibinfo {year} {1976})}\BibitemShut {NoStop}%
\bibitem [{\citenamefont {Timofeev}(2000)}]{Timofeev2000}%
  \BibitemOpen
  \bibfield  {author} {\bibinfo {author} {\bibfnamefont {A.~V.}\ \bibnamefont
  {Timofeev}},\ }\href {http://dx.doi.org/10.1134/1.952900} {\bibfield
  {journal} {\bibinfo  {journal} {Plasma Phys. Rep.}\ }\textbf {\bibinfo
  {volume} {26}},\ \bibinfo {pages} {626} (\bibinfo {year} {2000})}\BibitemShut
  {NoStop}%
\bibitem [{\citenamefont {Ochs}\ \emph
  {et~al.}(2017{\natexlab{a}})\citenamefont {Ochs}, \citenamefont {Rax},
  \citenamefont {Gueroult},\ and\ \citenamefont {Fisch}}]{Ochs2017a}%
  \BibitemOpen
  \bibfield  {author} {\bibinfo {author} {\bibfnamefont {I.~E.}\ \bibnamefont
  {Ochs}}, \bibinfo {author} {\bibfnamefont {J.-M.}\ \bibnamefont {Rax}},
  \bibinfo {author} {\bibfnamefont {R.}~\bibnamefont {Gueroult}}, \ and\
  \bibinfo {author} {\bibfnamefont {N.~J.}\ \bibnamefont {Fisch}},\ }\href
  {\doibase 10.1063/1.4994327} {\bibfield  {journal} {\bibinfo  {journal}
  {Phys. Plasmas}\ }\textbf {\bibinfo {volume} {24}},\ \bibinfo {pages}
  {083503} (\bibinfo {year} {2017}{\natexlab{a}})}\BibitemShut {NoStop}%
\bibitem [{\citenamefont {Soubbaramayer}(1979)}]{Soubbaramayer1979}%
  \BibitemOpen
  \bibfield  {author} {\bibinfo {author} {\bibnamefont {Soubbaramayer}},\
  }\enquote {\bibinfo {title} {Centrifugation},}\ in\ \href {\doibase
  10.1007/3-540-09385-0_18} {\emph {\bibinfo {booktitle} {Uranium
  Enrichment}}}\ (\bibinfo  {publisher} {Springer Berlin Heidelberg},\ \bibinfo
  {year} {1979})\ pp.\ \bibinfo {pages} {183--244}\BibitemShut {NoStop}%
\bibitem [{\citenamefont {Bonnevier}(1966)}]{Bonnevier1966}%
  \BibitemOpen
  \bibfield  {author} {\bibinfo {author} {\bibfnamefont {B.}~\bibnamefont
  {Bonnevier}},\ }\href@noop {} {\bibfield  {journal} {\bibinfo  {journal}
  {Ark. Fys.}\ }\textbf {\bibinfo {volume} {33}},\ \bibinfo {pages} {255}
  (\bibinfo {year} {1966})}\BibitemShut {NoStop}%
\bibitem [{\citenamefont {Lehnert}(1971)}]{Lehnert1971}%
  \BibitemOpen
  \bibfield  {author} {\bibinfo {author} {\bibfnamefont {B.}~\bibnamefont
  {Lehnert}},\ }\href {\doibase 10.1088/0029-5515/11/5/010} {\bibfield
  {journal} {\bibinfo  {journal} {Nucl. Fusion}\ }\textbf {\bibinfo {volume}
  {11}},\ \bibinfo {pages} {485} (\bibinfo {year} {1971})}\BibitemShut
  {NoStop}%
\bibitem [{\citenamefont {O'Neil}(1981)}]{Oneil1981}%
  \BibitemOpen
  \bibfield  {author} {\bibinfo {author} {\bibfnamefont {T.~M.}\ \bibnamefont
  {O'Neil}},\ }\href {\doibase 10.1063/1.863565} {\bibfield  {journal}
  {\bibinfo  {journal} {Phys.Fluids}\ }\textbf {\bibinfo {volume} {24}},\
  \bibinfo {pages} {1447} (\bibinfo {year} {1981})}\BibitemShut {NoStop}%
\bibitem [{\citenamefont {Bittencourt}\ and\ \citenamefont
  {Ludwig}(1987)}]{Bittencourt1987}%
  \BibitemOpen
  \bibfield  {author} {\bibinfo {author} {\bibfnamefont {J.~A.}\ \bibnamefont
  {Bittencourt}}\ and\ \bibinfo {author} {\bibfnamefont {G.~O.}\ \bibnamefont
  {Ludwig}},\ }\href {\doibase 10.1088/0741-3335/29/5/003} {\bibfield
  {journal} {\bibinfo  {journal} {Plasma Phys. Controlled Fusion}\ }\textbf
  {\bibinfo {volume} {29}},\ \bibinfo {pages} {601} (\bibinfo {year}
  {1987})}\BibitemShut {NoStop}%
\bibitem [{\citenamefont {Angerth}\ \emph {et~al.}(1962)\citenamefont
  {Angerth}, \citenamefont {Block}, \citenamefont {Fahleson},\ and\
  \citenamefont {Soop}}]{Angerth1962}%
  \BibitemOpen
  \bibfield  {author} {\bibinfo {author} {\bibfnamefont {B.}~\bibnamefont
  {Angerth}}, \bibinfo {author} {\bibfnamefont {L.}~\bibnamefont {Block}},
  \bibinfo {author} {\bibfnamefont {U.}~\bibnamefont {Fahleson}}, \ and\
  \bibinfo {author} {\bibfnamefont {C.}~\bibnamefont {Soop}},\ }\href@noop {}
  {\bibfield  {journal} {\bibinfo  {journal} {Nucl. Fusion}\ }\textbf {\bibinfo
  {volume} {Suppl, Part 1}},\ \bibinfo {pages} {39} (\bibinfo {year}
  {1962})}\BibitemShut {NoStop}%
\bibitem [{\citenamefont {Anderson}\ \emph {et~al.}(1958)\citenamefont
  {Anderson}, \citenamefont {Baker}, \citenamefont {Bratenahl}, \citenamefont
  {Ise}, \citenamefont {Kunkel}, \citenamefont {Stone},\ and\ \citenamefont
  {Furth}}]{Anderson1958}%
  \BibitemOpen
  \bibfield  {author} {\bibinfo {author} {\bibfnamefont {O.~A.}\ \bibnamefont
  {Anderson}}, \bibinfo {author} {\bibfnamefont {W.~R.}\ \bibnamefont {Baker}},
  \bibinfo {author} {\bibfnamefont {A.}~\bibnamefont {Bratenahl}}, \bibinfo
  {author} {\bibfnamefont {J.~J.}\ \bibnamefont {Ise}}, \bibinfo {author}
  {\bibfnamefont {W.~B.}\ \bibnamefont {Kunkel}}, \bibinfo {author}
  {\bibfnamefont {J.~M.}\ \bibnamefont {Stone}}, \ and\ \bibinfo {author}
  {\bibfnamefont {H.~P.}\ \bibnamefont {Furth}},\ }in\ \href@noop {} {\emph
  {\bibinfo {booktitle} {Proceedings of the United Nations international
  conference on the peaceful uses of atomic energy}}}\ (\bibinfo {year}
  {1958})\ pp.\ \bibinfo {pages} {155--160}\BibitemShut {NoStop}%
\bibitem [{\citenamefont {Ohkawa}\ and\ \citenamefont
  {Miller}(2002)}]{Ohkawa2002}%
  \BibitemOpen
  \bibfield  {author} {\bibinfo {author} {\bibfnamefont {T.}~\bibnamefont
  {Ohkawa}}\ and\ \bibinfo {author} {\bibfnamefont {R.~L.}\ \bibnamefont
  {Miller}},\ }\href {\doibase 10.1063/1.1523930} {\bibfield  {journal}
  {\bibinfo  {journal} {Phys. Plasmas}\ }\textbf {\bibinfo {volume} {9}},\
  \bibinfo {pages} {5116} (\bibinfo {year} {2002})}\BibitemShut {NoStop}%
\bibitem [{\citenamefont {Gueroult}, \citenamefont {Rax},\ and\ \citenamefont
  {Fisch}(2014)}]{Gueroult2014}%
  \BibitemOpen
  \bibfield  {author} {\bibinfo {author} {\bibfnamefont {R.}~\bibnamefont
  {Gueroult}}, \bibinfo {author} {\bibfnamefont {J.-M.}\ \bibnamefont {Rax}}, \
  and\ \bibinfo {author} {\bibfnamefont {N.~J.}\ \bibnamefont {Fisch}},\ }\href
  {\doibase 10.1063/1.4864325} {\bibfield  {journal} {\bibinfo  {journal}
  {Phys. Plasmas}\ }\textbf {\bibinfo {volume} {21}},\ \bibinfo {pages}
  {020701} (\bibinfo {year} {2014})}\BibitemShut {NoStop}%
\bibitem [{\citenamefont {Wijnakker}, \citenamefont {Granneman},\ and\
  \citenamefont {Kistemaker}(1979)}]{Wijnakker1979}%
  \BibitemOpen
  \bibfield  {author} {\bibinfo {author} {\bibfnamefont {M.~M.~B.}\
  \bibnamefont {Wijnakker}}, \bibinfo {author} {\bibfnamefont {E.~H.~A.}\
  \bibnamefont {Granneman}}, \ and\ \bibinfo {author} {\bibfnamefont
  {J.}~\bibnamefont {Kistemaker}},\ }\href@noop {} {\bibfield  {journal}
  {\bibinfo  {journal} {Z. Naturforsch.}\ }\textbf {\bibinfo {volume} {34a}},\
  \bibinfo {pages} {672} (\bibinfo {year} {1979})}\BibitemShut {NoStop}%
\bibitem [{\citenamefont {Barber}, \citenamefont {Swift},\ and\ \citenamefont
  {Tozer}(1972)}]{Barber1972}%
  \BibitemOpen
  \bibfield  {author} {\bibinfo {author} {\bibfnamefont {P.~B.}\ \bibnamefont
  {Barber}}, \bibinfo {author} {\bibfnamefont {D.~A.}\ \bibnamefont {Swift}}, \
  and\ \bibinfo {author} {\bibfnamefont {B.~A.}\ \bibnamefont {Tozer}},\ }\href
  {\doibase 10.1088/0022-3727/5/4/308} {\bibfield  {journal} {\bibinfo
  {journal} {J. Phys. D: Appl. Phys.}\ }\textbf {\bibinfo {volume} {5}},\
  \bibinfo {pages} {693} (\bibinfo {year} {1972})}\BibitemShut {NoStop}%
\bibitem [{\citenamefont {James}\ and\ \citenamefont
  {Simpson}(1976)}]{James1976}%
  \BibitemOpen
  \bibfield  {author} {\bibinfo {author} {\bibfnamefont {B.~W.}\ \bibnamefont
  {James}}\ and\ \bibinfo {author} {\bibfnamefont {S.~W.}\ \bibnamefont
  {Simpson}},\ }\href {\doibase 10.1088/0032-1028/18/4/004} {\bibfield
  {journal} {\bibinfo  {journal} {Plasma Phys.}\ }\textbf {\bibinfo {volume}
  {18}},\ \bibinfo {pages} {289} (\bibinfo {year} {1976})}\BibitemShut
  {NoStop}%
\bibitem [{\citenamefont {Wijnakker}\ and\ \citenamefont
  {Granneman}(1980)}]{Wijnakker1980}%
  \BibitemOpen
  \bibfield  {author} {\bibinfo {author} {\bibfnamefont {M.~M.~B.}\
  \bibnamefont {Wijnakker}}\ and\ \bibinfo {author} {\bibfnamefont {E.~H.~A.}\
  \bibnamefont {Granneman}},\ }\href@noop {} {\bibfield  {journal} {\bibinfo
  {journal} {Z. Naturforsch.}\ }\textbf {\bibinfo {volume} {35a}},\ \bibinfo
  {pages} {883} (\bibinfo {year} {1980})}\BibitemShut {NoStop}%
\bibitem [{\citenamefont {Krishnan}, \citenamefont {Geva},\ and\ \citenamefont
  {Hirshfield}(1981)}]{Krishnan1981}%
  \BibitemOpen
  \bibfield  {author} {\bibinfo {author} {\bibfnamefont {M.}~\bibnamefont
  {Krishnan}}, \bibinfo {author} {\bibfnamefont {M.}~\bibnamefont {Geva}}, \
  and\ \bibinfo {author} {\bibfnamefont {J.~L.}\ \bibnamefont {Hirshfield}},\
  }\href {\doibase 10.1103/PhysRevLett.46.36} {\bibfield  {journal} {\bibinfo
  {journal} {Phys. Rev. Lett.}\ }\textbf {\bibinfo {volume} {46}},\ \bibinfo
  {pages} {36} (\bibinfo {year} {1981})}\BibitemShut {NoStop}%
\bibitem [{\citenamefont {Alfv{\'e}n}(1960)}]{Alfven1960}%
  \BibitemOpen
  \bibfield  {author} {\bibinfo {author} {\bibfnamefont {H.}~\bibnamefont
  {Alfv{\'e}n}},\ }\href {https://dx.doi.org/10.1103/RevModPhys.32.710}
  {\bibfield  {journal} {\bibinfo  {journal} {Rev. Mod. Phys.}\ }\textbf
  {\bibinfo {volume} {32}},\ \bibinfo {pages} {710} (\bibinfo {year}
  {1960})}\BibitemShut {NoStop}%
\bibitem [{\citenamefont {Prasad}\ and\ \citenamefont
  {Krishnan}(1987{\natexlab{a}})}]{Prasad1987}%
  \BibitemOpen
  \bibfield  {author} {\bibinfo {author} {\bibfnamefont {R.~R.}\ \bibnamefont
  {Prasad}}\ and\ \bibinfo {author} {\bibfnamefont {M.}~\bibnamefont
  {Krishnan}},\ }\href {\doibase 10.1063/1.866430} {\bibfield  {journal}
  {\bibinfo  {journal} {Phys. Fluids}\ }\textbf {\bibinfo {volume} {30}},\
  \bibinfo {pages} {3496} (\bibinfo {year} {1987}{\natexlab{a}})}\BibitemShut
  {NoStop}%
\bibitem [{\citenamefont {Hirshfield}, \citenamefont {Levin},\ and\
  \citenamefont {Danziger}(1989)}]{Hirshfield1989}%
  \BibitemOpen
  \bibfield  {author} {\bibinfo {author} {\bibfnamefont {J.~L.}\ \bibnamefont
  {Hirshfield}}, \bibinfo {author} {\bibfnamefont {L.~A.}\ \bibnamefont
  {Levin}}, \ and\ \bibinfo {author} {\bibfnamefont {O.}~\bibnamefont
  {Danziger}},\ }\href {\doibase 10.1109/27.41184} {\bibfield  {journal}
  {\bibinfo  {journal} {IEEE Trans. Plasma Sci.}\ }\textbf {\bibinfo {volume}
  {17}},\ \bibinfo {pages} {695} (\bibinfo {year} {1989})}\BibitemShut
  {NoStop}%
\bibitem [{\citenamefont {Krishnan}(1983)}]{Krishnan1983}%
  \BibitemOpen
  \bibfield  {author} {\bibinfo {author} {\bibfnamefont {M.}~\bibnamefont
  {Krishnan}},\ }\href {\doibase 10.1063/1.864460} {\bibfield  {journal}
  {\bibinfo  {journal} {Phys. Fluids}\ }\textbf {\bibinfo {volume} {26}},\
  \bibinfo {pages} {2676} (\bibinfo {year} {1983})}\BibitemShut {NoStop}%
\bibitem [{\citenamefont {Poluektov}\ and\ \citenamefont
  {Efremov}(1998)}]{Poluektov1998}%
  \BibitemOpen
  \bibfield  {author} {\bibinfo {author} {\bibfnamefont {N.~P.}\ \bibnamefont
  {Poluektov}}\ and\ \bibinfo {author} {\bibfnamefont {N.~P.}\ \bibnamefont
  {Efremov}},\ }\href {\doibase 10.1088/0022-3727/31/8/010} {\bibfield
  {journal} {\bibinfo  {journal} {J. Phys. D: Appl. Phys.}\ }\textbf {\bibinfo
  {volume} {31}},\ \bibinfo {pages} {988} (\bibinfo {year} {1998})}\BibitemShut
  {NoStop}%
\bibitem [{\citenamefont {Hole}\ \emph {et~al.}(2002)\citenamefont {Hole},
  \citenamefont {Dallaqua}, \citenamefont {Simpson},\ and\ \citenamefont
  {Del~Bosco}}]{Hole2002}%
  \BibitemOpen
  \bibfield  {author} {\bibinfo {author} {\bibfnamefont {M.~J.}\ \bibnamefont
  {Hole}}, \bibinfo {author} {\bibfnamefont {R.~S.}\ \bibnamefont {Dallaqua}},
  \bibinfo {author} {\bibfnamefont {S.~W.}\ \bibnamefont {Simpson}}, \ and\
  \bibinfo {author} {\bibfnamefont {E.}~\bibnamefont {Del~Bosco}},\ }\href
  {\doibase 10.1103/PhysRevE.65.046409} {\bibfield  {journal} {\bibinfo
  {journal} {Phys. Rev. E}\ }\textbf {\bibinfo {volume} {65}},\ \bibinfo
  {pages} {046409} (\bibinfo {year} {2002})}\BibitemShut {NoStop}%
\bibitem [{\citenamefont {Prasad}\ and\ \citenamefont
  {Krishnan}(1987{\natexlab{b}})}]{Prasad1987b}%
  \BibitemOpen
  \bibfield  {author} {\bibinfo {author} {\bibfnamefont {R.~R.}\ \bibnamefont
  {Prasad}}\ and\ \bibinfo {author} {\bibfnamefont {M.}~\bibnamefont
  {Krishnan}},\ }\href {\doibase 10.1063/1.338976} {\bibfield  {journal}
  {\bibinfo  {journal} {J. Appl. Phys.}\ }\textbf {\bibinfo {volume} {61}},\
  \bibinfo {pages} {113} (\bibinfo {year} {1987}{\natexlab{b}})}\BibitemShut
  {NoStop}%
\bibitem [{\citenamefont {Prasad}\ and\ \citenamefont
  {Krishnan}(1987{\natexlab{c}})}]{Prasad1987a}%
  \BibitemOpen
  \bibfield  {author} {\bibinfo {author} {\bibfnamefont {R.~R.}\ \bibnamefont
  {Prasad}}\ and\ \bibinfo {author} {\bibfnamefont {M.}~\bibnamefont
  {Krishnan}},\ }\href
  {http://www.sciencedirect.com/science/article/pii/0168583X87907348}
  {\bibfield  {journal} {\bibinfo  {journal} {Nucl. Instrum. Methods Phys.
  Res., Sect. B}\ }\textbf {\bibinfo {volume} {26}},\ \bibinfo {pages} {65}
  (\bibinfo {year} {1987}{\natexlab{c}})}\BibitemShut {NoStop}%
\bibitem [{\citenamefont {Fetterman}\ and\ \citenamefont
  {Fisch}(2011{\natexlab{a}})}]{Fetterman2011b}%
  \BibitemOpen
  \bibfield  {author} {\bibinfo {author} {\bibfnamefont {A.~J.}\ \bibnamefont
  {Fetterman}}\ and\ \bibinfo {author} {\bibfnamefont {N.~J.}\ \bibnamefont
  {Fisch}},\ }\href {\doibase 10.1063/1.3646311} {\bibfield  {journal}
  {\bibinfo  {journal} {Phys. Plasmas}\ }\textbf {\bibinfo {volume} {18}},\
  \bibinfo {pages} {103503} (\bibinfo {year} {2011}{\natexlab{a}})}\BibitemShut
  {NoStop}%
\bibitem [{\citenamefont {Freeman}\ \emph {et~al.}(2003)\citenamefont
  {Freeman}, \citenamefont {Agnew}, \citenamefont {Anderegg}, \citenamefont
  {Cluggish}, \citenamefont {Gilleland}, \citenamefont {Isler}, \citenamefont
  {Litvak}, \citenamefont {Miller}, \citenamefont {O'Neill}, \citenamefont
  {Ohkawa}, \citenamefont {Pronko}, \citenamefont {Putvinski}, \citenamefont
  {Sevier}, \citenamefont {Sibley}, \citenamefont {Umstadter}, \citenamefont
  {Wade},\ and\ \citenamefont {Winslow}}]{Freeman2003}%
  \BibitemOpen
  \bibfield  {author} {\bibinfo {author} {\bibfnamefont {R.}~\bibnamefont
  {Freeman}}, \bibinfo {author} {\bibfnamefont {S.}~\bibnamefont {Agnew}},
  \bibinfo {author} {\bibfnamefont {F.}~\bibnamefont {Anderegg}}, \bibinfo
  {author} {\bibfnamefont {B.}~\bibnamefont {Cluggish}}, \bibinfo {author}
  {\bibfnamefont {J.}~\bibnamefont {Gilleland}}, \bibinfo {author}
  {\bibfnamefont {R.}~\bibnamefont {Isler}}, \bibinfo {author} {\bibfnamefont
  {A.}~\bibnamefont {Litvak}}, \bibinfo {author} {\bibfnamefont
  {R.}~\bibnamefont {Miller}}, \bibinfo {author} {\bibfnamefont
  {R.}~\bibnamefont {O'Neill}}, \bibinfo {author} {\bibfnamefont
  {T.}~\bibnamefont {Ohkawa}}, \bibinfo {author} {\bibfnamefont
  {S.}~\bibnamefont {Pronko}}, \bibinfo {author} {\bibfnamefont
  {S.}~\bibnamefont {Putvinski}}, \bibinfo {author} {\bibfnamefont
  {L.}~\bibnamefont {Sevier}}, \bibinfo {author} {\bibfnamefont
  {A.}~\bibnamefont {Sibley}}, \bibinfo {author} {\bibfnamefont
  {K.}~\bibnamefont {Umstadter}}, \bibinfo {author} {\bibfnamefont
  {T.}~\bibnamefont {Wade}}, \ and\ \bibinfo {author} {\bibfnamefont
  {D.}~\bibnamefont {Winslow}},\ }\href {\doibase 10.1063/1.1638067} {\bibfield
   {journal} {\bibinfo  {journal} {AIP Conf. Proc.}\ }\textbf {\bibinfo
  {volume} {694}},\ \bibinfo {pages} {403} (\bibinfo {year}
  {2003})}\BibitemShut {NoStop}%
\bibitem [{\citenamefont {Chen}(1966)}]{Chen1966}%
  \BibitemOpen
  \bibfield  {author} {\bibinfo {author} {\bibfnamefont {F.~F.}\ \bibnamefont
  {Chen}},\ }\href {\doibase 10.1063/1.1761798} {\bibfield  {journal} {\bibinfo
   {journal} {Phys. Fluids}\ }\textbf {\bibinfo {volume} {9}},\ \bibinfo
  {pages} {965} (\bibinfo {year} {1966})}\BibitemShut {NoStop}%
\bibitem [{\citenamefont {Gueroult}, \citenamefont {Rax},\ and\ \citenamefont
  {Fisch}(2017)}]{Gueroult2017b}%
  \BibitemOpen
  \bibfield  {author} {\bibinfo {author} {\bibfnamefont {R.}~\bibnamefont
  {Gueroult}}, \bibinfo {author} {\bibfnamefont {J.~M.}\ \bibnamefont {Rax}}, \
  and\ \bibinfo {author} {\bibfnamefont {N.~J.}\ \bibnamefont {Fisch}},\ }\href
  {\doibase 10.1063/1.4994546} {\bibfield  {journal} {\bibinfo  {journal}
  {Phys. Plasmas}\ }\textbf {\bibinfo {volume} {24}},\ \bibinfo {pages}
  {082102} (\bibinfo {year} {2017})}\BibitemShut {NoStop}%
\bibitem [{\citenamefont {Brenning}(1992)}]{Brenning1992}%
  \BibitemOpen
  \bibfield  {author} {\bibinfo {author} {\bibfnamefont {N.}~\bibnamefont
  {Brenning}},\ }\href {\doibase 10.1007/BF00242088} {\bibfield  {journal}
  {\bibinfo  {journal} {Space Sci. Rev.}\ }\textbf {\bibinfo {volume} {59}},\
  \bibinfo {pages} {209} (\bibinfo {year} {1992})}\BibitemShut {NoStop}%
\bibitem [{\citenamefont {Lai}(2001)}]{LAI2001}%
  \BibitemOpen
  \bibfield  {author} {\bibinfo {author} {\bibfnamefont {S.~T.}\ \bibnamefont
  {Lai}},\ }\href {\doibase 10.1029/2000RG000087} {\bibfield  {journal}
  {\bibinfo  {journal} {Rev. Geophys.}\ }\textbf {\bibinfo {volume} {39}},\
  \bibinfo {pages} {471} (\bibinfo {year} {2001})}\BibitemShut {NoStop}%
\bibitem [{\citenamefont {Shinohara}\ and\ \citenamefont
  {Horii}(2007)}]{Shinohara2007}%
  \BibitemOpen
  \bibfield  {author} {\bibinfo {author} {\bibfnamefont {S.}~\bibnamefont
  {Shinohara}}\ and\ \bibinfo {author} {\bibfnamefont {S.}~\bibnamefont
  {Horii}},\ }\href {\doibase 10.1143/JJAP.46.4276} {\bibfield  {journal}
  {\bibinfo  {journal} {Jap. J. App. Phys.}\ }\textbf {\bibinfo {volume}
  {46}},\ \bibinfo {pages} {4276} (\bibinfo {year} {2007})}\BibitemShut
  {NoStop}%
\bibitem [{\citenamefont {Fetterman}\ and\ \citenamefont
  {Fisch}(2011{\natexlab{b}})}]{Fetterman2011}%
  \BibitemOpen
  \bibfield  {author} {\bibinfo {author} {\bibfnamefont {A.~J.}\ \bibnamefont
  {Fetterman}}\ and\ \bibinfo {author} {\bibfnamefont {N.~J.}\ \bibnamefont
  {Fisch}},\ }\href {\doibase 10.1063/1.3631793} {\bibfield  {journal}
  {\bibinfo  {journal} {Phys. Plasmas}\ }\textbf {\bibinfo {volume} {18}},\
  \bibinfo {pages} {094503} (\bibinfo {year} {2011}{\natexlab{b}})}\BibitemShut
  {NoStop}%
\bibitem [{\citenamefont {Volosov}(1997)}]{Volosov1997}%
  \BibitemOpen
  \bibfield  {author} {\bibinfo {author} {\bibfnamefont {V.~I.}\ \bibnamefont
  {Volosov}},\ }\href@noop {} {\bibfield  {journal} {\bibinfo  {journal}
  {Plasma Phys. Rep.}\ }\textbf {\bibinfo {volume} {23}},\ \bibinfo {pages}
  {751} (\bibinfo {year} {1997})}\BibitemShut {NoStop}%
\bibitem [{\citenamefont {Volosov}(2006)}]{Volosov2006}%
  \BibitemOpen
  \bibfield  {author} {\bibinfo {author} {\bibfnamefont {V.}~\bibnamefont
  {Volosov}},\ }\href {\doibase 10.1088/0029-5515/46/8/007} {\bibfield
  {journal} {\bibinfo  {journal} {Nucl. Fusion}\ }\textbf {\bibinfo {volume}
  {46}},\ \bibinfo {pages} {820} (\bibinfo {year} {2006})}\BibitemShut
  {NoStop}%
\bibitem [{\citenamefont {Lehnert}(1974)}]{Lehnert1974}%
  \BibitemOpen
  \bibfield  {author} {\bibinfo {author} {\bibfnamefont {B.}~\bibnamefont
  {Lehnert}},\ }\href {\doibase 10.1088/0031-8949/9/3/009} {\bibfield
  {journal} {\bibinfo  {journal} {Phys. Scr.}\ }\textbf {\bibinfo {volume}
  {9}},\ \bibinfo {pages} {189} (\bibinfo {year} {1974})}\BibitemShut {NoStop}%
\bibitem [{\citenamefont {Bekhtenev}\ \emph {et~al.}(1980)\citenamefont
  {Bekhtenev}, \citenamefont {Volosov}, \citenamefont {Pal{'}chikov},
  \citenamefont {Pekker},\ and\ \citenamefont {Yudin}}]{Bekhtenev1980}%
  \BibitemOpen
  \bibfield  {author} {\bibinfo {author} {\bibfnamefont {A.~A.}\ \bibnamefont
  {Bekhtenev}}, \bibinfo {author} {\bibfnamefont {V.~I.}\ \bibnamefont
  {Volosov}}, \bibinfo {author} {\bibfnamefont {V.~E.}\ \bibnamefont
  {Pal{'}chikov}}, \bibinfo {author} {\bibfnamefont {M.~S.}\ \bibnamefont
  {Pekker}}, \ and\ \bibinfo {author} {\bibfnamefont {Y.~N.}\ \bibnamefont
  {Yudin}},\ }\href {\doibase 10.1088/0029-5515/20/5/007} {\bibfield  {journal}
  {\bibinfo  {journal} {Nucl. Fusion}\ }\textbf {\bibinfo {volume} {20}},\
  \bibinfo {pages} {579} (\bibinfo {year} {1980})}\BibitemShut {NoStop}%
\bibitem [{\citenamefont {Gueroult}\ and\ \citenamefont
  {Fisch}(2012)}]{Gueroult2012a}%
  \BibitemOpen
  \bibfield  {author} {\bibinfo {author} {\bibfnamefont {R.}~\bibnamefont
  {Gueroult}}\ and\ \bibinfo {author} {\bibfnamefont {N.~J.}\ \bibnamefont
  {Fisch}},\ }\href {\doibase 10.1063/1.4771674} {\bibfield  {journal}
  {\bibinfo  {journal} {Phys. Plasmas}\ }\textbf {\bibinfo {volume} {19}},\
  \bibinfo {pages} {122503} (\bibinfo {year} {2012})}\BibitemShut {NoStop}%
\bibitem [{\citenamefont {Ochs}\ \emph
  {et~al.}(2017{\natexlab{b}})\citenamefont {Ochs}, \citenamefont {Gueroult},
  \citenamefont {Fisch},\ and\ \citenamefont {Zweben}}]{Ochs2017}%
  \BibitemOpen
  \bibfield  {author} {\bibinfo {author} {\bibfnamefont {I.~E.}\ \bibnamefont
  {Ochs}}, \bibinfo {author} {\bibfnamefont {R.}~\bibnamefont {Gueroult}},
  \bibinfo {author} {\bibfnamefont {N.~J.}\ \bibnamefont {Fisch}}, \ and\
  \bibinfo {author} {\bibfnamefont {S.~J.}\ \bibnamefont {Zweben}},\ }\href
  {\doibase 10.1063/1.4978949} {\bibfield  {journal} {\bibinfo  {journal}
  {Phys. Plasmas}\ }\textbf {\bibinfo {volume} {24}},\ \bibinfo {pages}
  {043503} (\bibinfo {year} {2017}{\natexlab{b}})}\BibitemShut {NoStop}%
\bibitem [{\citenamefont {Abolmasov}(2012)}]{Abolmasov2012}%
  \BibitemOpen
  \bibfield  {author} {\bibinfo {author} {\bibfnamefont {S.~N.}\ \bibnamefont
  {Abolmasov}},\ }\href {\doibase 10.1088/0963-0252/21/3/035006} {\bibfield
  {journal} {\bibinfo  {journal} {Plasma Sources Sci. Technol.}\ }\textbf
  {\bibinfo {volume} {21}},\ \bibinfo {pages} {035006} (\bibinfo {year}
  {2012})}\BibitemShut {NoStop}%
\bibitem [{\citenamefont {Boeuf}(2017)}]{Boeuf2017}%
  \BibitemOpen
  \bibfield  {author} {\bibinfo {author} {\bibfnamefont {J.-P.}\ \bibnamefont
  {Boeuf}},\ }\href {\doibase 10.1063/1.4972269} {\bibfield  {journal}
  {\bibinfo  {journal} {J. Appl. Phys.}\ }\textbf {\bibinfo {volume} {121}},\
  \bibinfo {pages} {011101} (\bibinfo {year} {2017})}\BibitemShut {NoStop}%
\bibitem [{\citenamefont {Morozov}\ and\ \citenamefont
  {Savel{'}ev}(2005)}]{Morozov2005}%
  \BibitemOpen
  \bibfield  {author} {\bibinfo {author} {\bibfnamefont {A.~I.}\ \bibnamefont
  {Morozov}}\ and\ \bibinfo {author} {\bibfnamefont {V.~V.}\ \bibnamefont
  {Savel{'}ev}},\ }\href {http://dx.doi.org/10.1134/1.1925791} {\bibfield
  {journal} {\bibinfo  {journal} {Plasma Phys. Rep.}\ }\textbf {\bibinfo
  {volume} {31}},\ \bibinfo {pages} {417} (\bibinfo {year} {2005})}\BibitemShut
  {NoStop}%
\bibitem [{\citenamefont {Bardakov}\ \emph {et~al.}(2010)\citenamefont
  {Bardakov}, \citenamefont {Kichigin}, \citenamefont {Strokin},\ and\
  \citenamefont {Tsaregorodtsev}}]{Bardakov2010a}%
  \BibitemOpen
  \bibfield  {author} {\bibinfo {author} {\bibfnamefont {V.~M.}\ \bibnamefont
  {Bardakov}}, \bibinfo {author} {\bibfnamefont {G.~N.}\ \bibnamefont
  {Kichigin}}, \bibinfo {author} {\bibfnamefont {N.~A.}\ \bibnamefont
  {Strokin}}, \ and\ \bibinfo {author} {\bibfnamefont {E.~O.}\ \bibnamefont
  {Tsaregorodtsev}},\ }\href {\doibase 10.1134/s1063784210100178} {\bibfield
  {journal} {\bibinfo  {journal} {Tech. Phys.}\ }\textbf {\bibinfo {volume}
  {55}},\ \bibinfo {pages} {1504} (\bibinfo {year} {2010})}\BibitemShut
  {NoStop}%
\bibitem [{\citenamefont {Bardakov}, \citenamefont {Ivanov},\ and\
  \citenamefont {Strokin}(2014)}]{Bardakov2014}%
  \BibitemOpen
  \bibfield  {author} {\bibinfo {author} {\bibfnamefont {V.~M.}\ \bibnamefont
  {Bardakov}}, \bibinfo {author} {\bibfnamefont {S.~D.}\ \bibnamefont
  {Ivanov}}, \ and\ \bibinfo {author} {\bibfnamefont {N.~A.}\ \bibnamefont
  {Strokin}},\ }\href {\doibase 10.1063/1.4846898} {\bibfield  {journal}
  {\bibinfo  {journal} {Phys. Plasmas}\ }\textbf {\bibinfo {volume} {21}},\
  \bibinfo {pages} {033505} (\bibinfo {year} {2014})}\BibitemShut {NoStop}%
\bibitem [{\citenamefont {Smirnov}\ \emph {et~al.}(2013)\citenamefont
  {Smirnov}, \citenamefont {Samokhin}, \citenamefont {Vorona},\ and\
  \citenamefont {Gavrikov}}]{Smirnov2013}%
  \BibitemOpen
  \bibfield  {author} {\bibinfo {author} {\bibfnamefont {V.~P.}\ \bibnamefont
  {Smirnov}}, \bibinfo {author} {\bibfnamefont {A.~A.}\ \bibnamefont
  {Samokhin}}, \bibinfo {author} {\bibfnamefont {N.~A.}\ \bibnamefont
  {Vorona}}, \ and\ \bibinfo {author} {\bibfnamefont {A.~V.}\ \bibnamefont
  {Gavrikov}},\ }\href {http://dx.doi.org/10.1134/S1063780X13050103} {\bibfield
   {journal} {\bibinfo  {journal} {Plasma Phys. Rep.}\ }\textbf {\bibinfo
  {volume} {39}},\ \bibinfo {pages} {456} (\bibinfo {year} {2013})}\BibitemShut
  {NoStop}%
\bibitem [{\citenamefont {Bardakov}, \citenamefont {Kichigin},\ and\
  \citenamefont {Strokin}(2010)}]{Bardakov2010}%
  \BibitemOpen
  \bibfield  {author} {\bibinfo {author} {\bibfnamefont {V.~M.}\ \bibnamefont
  {Bardakov}}, \bibinfo {author} {\bibfnamefont {G.~N.}\ \bibnamefont
  {Kichigin}}, \ and\ \bibinfo {author} {\bibfnamefont {N.~A.}\ \bibnamefont
  {Strokin}},\ }\href {http://dx.doi.org/10.1134/S1063785010020276} {\bibfield
  {journal} {\bibinfo  {journal} {Tech. Phys. Lett.}\ }\textbf {\bibinfo
  {volume} {36}},\ \bibinfo {pages} {185} (\bibinfo {year} {2010})}\BibitemShut
  {NoStop}%
\bibitem [{\citenamefont {Bardakov}\ \emph {et~al.}(2015)\citenamefont
  {Bardakov}, \citenamefont {Ivanov}, \citenamefont {Kazantsev},\ and\
  \citenamefont {Strokin}}]{Bardakov2015}%
  \BibitemOpen
  \bibfield  {author} {\bibinfo {author} {\bibfnamefont {V.~M.}\ \bibnamefont
  {Bardakov}}, \bibinfo {author} {\bibfnamefont {S.~D.}\ \bibnamefont
  {Ivanov}}, \bibinfo {author} {\bibfnamefont {A.~V.}\ \bibnamefont
  {Kazantsev}}, \ and\ \bibinfo {author} {\bibfnamefont {N.~A.}\ \bibnamefont
  {Strokin}},\ }\href {\doibase 10.1088/1009-0630/17/10/09} {\bibfield
  {journal} {\bibinfo  {journal} {Plasma Sci. Technol}\ }\textbf {\bibinfo
  {volume} {17}},\ \bibinfo {pages} {862} (\bibinfo {year} {2015})}\BibitemShut
  {NoStop}%
\bibitem [{\citenamefont {Bardakov}\ \emph {et~al.}(2018)\citenamefont
  {Bardakov}, \citenamefont {Ivanov}, \citenamefont {Kazantsev}, \citenamefont
  {Strokin},\ and\ \citenamefont {Stupin}}]{Bardakov2018}%
  \BibitemOpen
  \bibfield  {author} {\bibinfo {author} {\bibfnamefont {V.~M.}\ \bibnamefont
  {Bardakov}}, \bibinfo {author} {\bibfnamefont {S.~D.}\ \bibnamefont
  {Ivanov}}, \bibinfo {author} {\bibfnamefont {A.~V.}\ \bibnamefont
  {Kazantsev}}, \bibinfo {author} {\bibfnamefont {N.~A.}\ \bibnamefont
  {Strokin}}, \ and\ \bibinfo {author} {\bibfnamefont {A.~N.}\ \bibnamefont
  {Stupin}},\ }\href {\doibase 10.1063/1.5037852} {\bibfield  {journal}
  {\bibinfo  {journal} {Phys. Plasmas}\ }\textbf {\bibinfo {volume} {25}},\
  \bibinfo {pages} {083509} (\bibinfo {year} {2018})}\BibitemShut {NoStop}%
\bibitem [{\citenamefont {Samokhin}\ \emph {et~al.}(2016)\citenamefont
  {Samokhin}, \citenamefont {Smirnov}, \citenamefont {Gavrikov},\ and\
  \citenamefont {Vorona}}]{Samokhin2016}%
  \BibitemOpen
  \bibfield  {author} {\bibinfo {author} {\bibfnamefont {A.~A.}\ \bibnamefont
  {Samokhin}}, \bibinfo {author} {\bibfnamefont {V.~P.}\ \bibnamefont
  {Smirnov}}, \bibinfo {author} {\bibfnamefont {A.~V.}\ \bibnamefont
  {Gavrikov}}, \ and\ \bibinfo {author} {\bibfnamefont {N.~A.}\ \bibnamefont
  {Vorona}},\ }\href {\doibase 10.1134/S1063784216020298} {\bibfield  {journal}
  {\bibinfo  {journal} {Tech. Phys.}\ }\textbf {\bibinfo {volume} {61}},\
  \bibinfo {pages} {283} (\bibinfo {year} {2016})}\BibitemShut {NoStop}%
\bibitem [{\citenamefont {Bagryansky}\ \emph {et~al.}(2003)\citenamefont
  {Bagryansky}, \citenamefont {Lizunov}, \citenamefont {Zuev}, \citenamefont
  {Kolesnikov},\ and\ \citenamefont {Solomachin}}]{Bagryansky2003}%
  \BibitemOpen
  \bibfield  {author} {\bibinfo {author} {\bibfnamefont {P.~A.}\ \bibnamefont
  {Bagryansky}}, \bibinfo {author} {\bibfnamefont {A.~A.}\ \bibnamefont
  {Lizunov}}, \bibinfo {author} {\bibfnamefont {A.~A.}\ \bibnamefont {Zuev}},
  \bibinfo {author} {\bibfnamefont {E.~Y.}\ \bibnamefont {Kolesnikov}}, \ and\
  \bibinfo {author} {\bibfnamefont {A.~L.}\ \bibnamefont {Solomachin}},\
  }\href@noop {} {\bibfield  {journal} {\bibinfo  {journal} {Transactions of
  Fusion Science and Technology}\ }\textbf {\bibinfo {volume} {43}},\ \bibinfo
  {pages} {152} (\bibinfo {year} {2003})}\BibitemShut {NoStop}%
\bibitem [{\citenamefont {Schaffner}\ \emph {et~al.}(2012)\citenamefont
  {Schaffner}, \citenamefont {Carter}, \citenamefont {Rossi}, \citenamefont
  {Guice}, \citenamefont {Maggs}, \citenamefont {Vincena},\ and\ \citenamefont
  {Friedman}}]{Schaffner2012}%
  \BibitemOpen
  \bibfield  {author} {\bibinfo {author} {\bibfnamefont {D.~A.}\ \bibnamefont
  {Schaffner}}, \bibinfo {author} {\bibfnamefont {T.~A.}\ \bibnamefont
  {Carter}}, \bibinfo {author} {\bibfnamefont {G.~D.}\ \bibnamefont {Rossi}},
  \bibinfo {author} {\bibfnamefont {D.~S.}\ \bibnamefont {Guice}}, \bibinfo
  {author} {\bibfnamefont {J.~E.}\ \bibnamefont {Maggs}}, \bibinfo {author}
  {\bibfnamefont {S.}~\bibnamefont {Vincena}}, \ and\ \bibinfo {author}
  {\bibfnamefont {B.}~\bibnamefont {Friedman}},\ }\href {\doibase
  10.1103/PhysRevLett.109.135002} {\bibfield  {journal} {\bibinfo  {journal}
  {Phys. Rev. Lett.}\ }\textbf {\bibinfo {volume} {109}},\ \bibinfo {pages}
  {135002} (\bibinfo {year} {2012})}\BibitemShut {NoStop}%
\bibitem [{\citenamefont {Weynants}\ and\ \citenamefont
  {Oost}(1993)}]{Weynants1993}%
  \BibitemOpen
  \bibfield  {author} {\bibinfo {author} {\bibfnamefont {R.~R.}\ \bibnamefont
  {Weynants}}\ and\ \bibinfo {author} {\bibfnamefont {G.~V.}\ \bibnamefont
  {Oost}},\ }\href {\doibase 10.1088/0741-3335/35/SB/014} {\bibfield  {journal}
  {\bibinfo  {journal} {Plasma Phys. Controlled Fusion}\ }\textbf {\bibinfo
  {volume} {35}},\ \bibinfo {pages} {B177} (\bibinfo {year}
  {1993})}\BibitemShut {NoStop}%
\bibitem [{\citenamefont {Lehnert}(1970)}]{Lehnert1970}%
  \BibitemOpen
  \bibfield  {author} {\bibinfo {author} {\bibfnamefont {B.}~\bibnamefont
  {Lehnert}},\ }\href {\doibase 10.1088/0031-8949/2/3/007} {\bibfield
  {journal} {\bibinfo  {journal} {Phys. Scr.}\ }\textbf {\bibinfo {volume}
  {2}},\ \bibinfo {pages} {106} (\bibinfo {year} {1970})}\BibitemShut {NoStop}%
\bibitem [{\citenamefont {Lehnert}(1973)}]{Lehnert1973}%
  \BibitemOpen
  \bibfield  {author} {\bibinfo {author} {\bibfnamefont {B.}~\bibnamefont
  {Lehnert}},\ }\href {\doibase 10.1088/0031-8949/7/3/002} {\bibfield
  {journal} {\bibinfo  {journal} {Phys. Scr.}\ }\textbf {\bibinfo {volume}
  {7}},\ \bibinfo {pages} {102} (\bibinfo {year} {1973})}\BibitemShut {NoStop}%
\bibitem [{\citenamefont {Bekhtenev}\ and\ \citenamefont
  {Volosov}(1978)}]{Bekhtenev1978}%
  \BibitemOpen
  \bibfield  {author} {\bibinfo {author} {\bibfnamefont {A.~A.}\ \bibnamefont
  {Bekhtenev}}\ and\ \bibinfo {author} {\bibfnamefont {V.}~\bibnamefont
  {Volosov}},\ }\href@noop {} {\bibfield  {journal} {\bibinfo  {journal}
  {Soviet Physics: Technical Physics}\ }\textbf {\bibinfo {volume} {23}},\
  \bibinfo {pages} {938} (\bibinfo {year} {1978})}\BibitemShut {NoStop}%
\bibitem [{\citenamefont {Helander}\ and\ \citenamefont
  {Sigmar}(2005)}]{Helander2005}%
  \BibitemOpen
  \bibfield  {author} {\bibinfo {author} {\bibfnamefont {P.}~\bibnamefont
  {Helander}}\ and\ \bibinfo {author} {\bibfnamefont {D.~J.}\ \bibnamefont
  {Sigmar}},\ }\href@noop {} {\emph {\bibinfo {title} {Collisional Transport in
  Magnetized Plasmas}}}\ (\bibinfo  {publisher} {Cambridge University Press},\
  \bibinfo {year} {2005})\BibitemShut {NoStop}%
\bibitem [{\citenamefont {Rozhansky}(2008)}]{Rozhansky2008}%
  \BibitemOpen
  \bibfield  {author} {\bibinfo {author} {\bibfnamefont {V.}~\bibnamefont
  {Rozhansky}},\ }in\ \href {\doibase 10.1007/978-3-540-74576-1} {\emph
  {\bibinfo {booktitle} {Reviews of Plasma Physics}}},\ Vol.~\bibinfo {volume}
  {24},\ \bibinfo {editor} {edited by\ \bibinfo {editor} {\bibfnamefont
  {V.~D.}\ \bibnamefont {Shafranov}}}\ (\bibinfo  {publisher} {Springer-Verlag
  Berlin Heidelberg},\ \bibinfo {year} {2008})\BibitemShut {NoStop}%
\bibitem [{\citenamefont {Rax}\ \emph {et~al.}(2018)\citenamefont {Rax},
  \citenamefont {Kolmes}, \citenamefont {Ochs}, \citenamefont {Fisch},\ and\
  \citenamefont {Gueroult}}]{Rax2018a}%
  \BibitemOpen
  \bibfield  {author} {\bibinfo {author} {\bibfnamefont {J.-M.}\ \bibnamefont
  {Rax}}, \bibinfo {author} {\bibfnamefont {E.~J.}\ \bibnamefont {Kolmes}},
  \bibinfo {author} {\bibfnamefont {I.~E.}\ \bibnamefont {Ochs}}, \bibinfo
  {author} {\bibfnamefont {N.~J.}\ \bibnamefont {Fisch}}, \ and\ \bibinfo
  {author} {\bibfnamefont {R.}~\bibnamefont {Gueroult}},\ }\href
  {https://arxiv.org/abs/1810.03696} {\bibfield  {journal} {\bibinfo  {journal}
  {ArXiv e-prints}\ } (\bibinfo {year} {2018})},\ \bibinfo {note}
  {1810.03696}\BibitemShut {NoStop}%
\bibitem [{\citenamefont {Horton}(1999)}]{Horton1999}%
  \BibitemOpen
  \bibfield  {author} {\bibinfo {author} {\bibfnamefont {W.}~\bibnamefont
  {Horton}},\ }\href {\doibase 10.1103/RevModPhys.71.735} {\bibfield  {journal}
  {\bibinfo  {journal} {Rev. Mod. Phys.}\ }\textbf {\bibinfo {volume} {71}},\
  \bibinfo {pages} {735} (\bibinfo {year} {1999})}\BibitemShut {NoStop}%
\bibitem [{\citenamefont {Jassby}(1972)}]{Jassby1972}%
  \BibitemOpen
  \bibfield  {author} {\bibinfo {author} {\bibfnamefont {D.~L.}\ \bibnamefont
  {Jassby}},\ }\href {\doibase 10.1063/1.1694135} {\bibfield  {journal}
  {\bibinfo  {journal} {Phys. Fluids}\ }\textbf {\bibinfo {volume} {15}},\
  \bibinfo {pages} {1590} (\bibinfo {year} {1972})}\BibitemShut {NoStop}%
\bibitem [{\citenamefont {Komori}, \citenamefont {Watanabe},\ and\
  \citenamefont {Kawai}(1988)}]{Komori1988}%
  \BibitemOpen
  \bibfield  {author} {\bibinfo {author} {\bibfnamefont {A.}~\bibnamefont
  {Komori}}, \bibinfo {author} {\bibfnamefont {K.}~\bibnamefont {Watanabe}}, \
  and\ \bibinfo {author} {\bibfnamefont {Y.}~\bibnamefont {Kawai}},\ }\href
  {\doibase 10.1063/1.866569} {\bibfield  {journal} {\bibinfo  {journal} {Phys.
  Fluids}\ }\textbf {\bibinfo {volume} {31}},\ \bibinfo {pages} {210} (\bibinfo
  {year} {1988})}\BibitemShut {NoStop}%
\bibitem [{\citenamefont {Amatucci}\ \emph {et~al.}(1996)\citenamefont
  {Amatucci}, \citenamefont {Walker}, \citenamefont {Ganguli}, \citenamefont
  {Antoniades}, \citenamefont {Duncan}, \citenamefont {Bowles}, \citenamefont
  {Gavrishchaka},\ and\ \citenamefont {Koepke}}]{Amatucci1996}%
  \BibitemOpen
  \bibfield  {author} {\bibinfo {author} {\bibfnamefont {W.~E.}\ \bibnamefont
  {Amatucci}}, \bibinfo {author} {\bibfnamefont {D.~N.}\ \bibnamefont
  {Walker}}, \bibinfo {author} {\bibfnamefont {G.}~\bibnamefont {Ganguli}},
  \bibinfo {author} {\bibfnamefont {J.~A.}\ \bibnamefont {Antoniades}},
  \bibinfo {author} {\bibfnamefont {D.}~\bibnamefont {Duncan}}, \bibinfo
  {author} {\bibfnamefont {J.~H.}\ \bibnamefont {Bowles}}, \bibinfo {author}
  {\bibfnamefont {V.}~\bibnamefont {Gavrishchaka}}, \ and\ \bibinfo {author}
  {\bibfnamefont {M.~E.}\ \bibnamefont {Koepke}},\ }\href
  {http://link.aps.org/doi/10.1103/PhysRevLett.77.1978} {\bibfield  {journal}
  {\bibinfo  {journal} {Phys. Rev. Lett.}\ }\textbf {\bibinfo {volume} {77}},\
  \bibinfo {pages} {1978} (\bibinfo {year} {1996})}\BibitemShut {NoStop}%
\bibitem [{\citenamefont {Mase}\ \emph {et~al.}(1991)\citenamefont {Mase},
  \citenamefont {Itakura}, \citenamefont {Inutake}, \citenamefont {Ishii},
  \citenamefont {Jeong}, \citenamefont {Hattori},\ and\ \citenamefont
  {Miyoshi}}]{Mase1991}%
  \BibitemOpen
  \bibfield  {author} {\bibinfo {author} {\bibfnamefont {A.}~\bibnamefont
  {Mase}}, \bibinfo {author} {\bibfnamefont {A.}~\bibnamefont {Itakura}},
  \bibinfo {author} {\bibfnamefont {M.}~\bibnamefont {Inutake}}, \bibinfo
  {author} {\bibfnamefont {K.}~\bibnamefont {Ishii}}, \bibinfo {author}
  {\bibfnamefont {J.~H.}\ \bibnamefont {Jeong}}, \bibinfo {author}
  {\bibfnamefont {K.}~\bibnamefont {Hattori}}, \ and\ \bibinfo {author}
  {\bibfnamefont {S.}~\bibnamefont {Miyoshi}},\ }\href
  {http://stacks.iop.org/0029-5515/31/i=9/a=010} {\bibfield  {journal}
  {\bibinfo  {journal} {Nucl. Fusion}\ }\textbf {\bibinfo {volume} {31}},\
  \bibinfo {pages} {1725} (\bibinfo {year} {1991})}\BibitemShut {NoStop}%
\bibitem [{\citenamefont {Abdrashitov}\ \emph {et~al.}(1991)\citenamefont
  {Abdrashitov}, \citenamefont {Beloborodov}, \citenamefont {Volosov},
  \citenamefont {Kubarev}, \citenamefont {Popov},\ and\ \citenamefont
  {Yudin}}]{Abdrashitov1991}%
  \BibitemOpen
  \bibfield  {author} {\bibinfo {author} {\bibfnamefont {G.~F.}\ \bibnamefont
  {Abdrashitov}}, \bibinfo {author} {\bibfnamefont {A.~V.}\ \bibnamefont
  {Beloborodov}}, \bibinfo {author} {\bibfnamefont {V.~I.}\ \bibnamefont
  {Volosov}}, \bibinfo {author} {\bibfnamefont {V.~V.}\ \bibnamefont
  {Kubarev}}, \bibinfo {author} {\bibfnamefont {Y.~S.}\ \bibnamefont {Popov}},
  \ and\ \bibinfo {author} {\bibfnamefont {Y.~N.}\ \bibnamefont {Yudin}},\
  }\href {\doibase 10.1088/0029-5515/31/7/004} {\bibfield  {journal} {\bibinfo
  {journal} {Nucl. Fusion}\ }\textbf {\bibinfo {volume} {31}},\ \bibinfo
  {pages} {1275} (\bibinfo {year} {1991})}\BibitemShut {NoStop}%
\bibitem [{\citenamefont {Severn}\ and\ \citenamefont
  {Hershkowitz}(1992)}]{Severn1992}%
  \BibitemOpen
  \bibfield  {author} {\bibinfo {author} {\bibfnamefont {G.~D.}\ \bibnamefont
  {Severn}}\ and\ \bibinfo {author} {\bibfnamefont {N.}~\bibnamefont
  {Hershkowitz}},\ }\href {\doibase 10.1063/1.860427} {\bibfield  {journal}
  {\bibinfo  {journal} {Phys. Fluids B: Plasma Phys.}\ }\textbf {\bibinfo
  {volume} {4}},\ \bibinfo {pages} {3210} (\bibinfo {year} {1992})}\BibitemShut
  {NoStop}%
\bibitem [{\citenamefont {Tsushima}\ \emph {et~al.}(1986)\citenamefont
  {Tsushima}, \citenamefont {Mieno}, \citenamefont {Oertl}, \citenamefont
  {Hatakeyama},\ and\ \citenamefont {Sato}}]{Tsushima1986}%
  \BibitemOpen
  \bibfield  {author} {\bibinfo {author} {\bibfnamefont {A.}~\bibnamefont
  {Tsushima}}, \bibinfo {author} {\bibfnamefont {T.}~\bibnamefont {Mieno}},
  \bibinfo {author} {\bibfnamefont {M.}~\bibnamefont {Oertl}}, \bibinfo
  {author} {\bibfnamefont {R.}~\bibnamefont {Hatakeyama}}, \ and\ \bibinfo
  {author} {\bibfnamefont {N.}~\bibnamefont {Sato}},\ }\href {\doibase
  10.1103/PhysRevLett.56.1815} {\bibfield  {journal} {\bibinfo  {journal}
  {Phys. Rev. Lett.}\ }\textbf {\bibinfo {volume} {56}},\ \bibinfo {pages}
  {1815} (\bibinfo {year} {1986})}\BibitemShut {NoStop}%
\bibitem [{\citenamefont {Tsushima}\ and\ \citenamefont
  {Sato}(1991)}]{Tsushima1991}%
  \BibitemOpen
  \bibfield  {author} {\bibinfo {author} {\bibfnamefont {A.}~\bibnamefont
  {Tsushima}}\ and\ \bibinfo {author} {\bibfnamefont {N.}~\bibnamefont
  {Sato}},\ }\bibfield  {booktitle} {\emph {\bibinfo {booktitle} {Journal of
  the Physical Society of Japan}},\ }\href {\doibase 10.1143/JPSJ.60.2665}
  {\bibfield  {journal} {\bibinfo  {journal} {J. Phys. Soc. Jpn.}\ }\textbf
  {\bibinfo {volume} {60}},\ \bibinfo {pages} {2665} (\bibinfo {year}
  {1991})}\BibitemShut {NoStop}%
\bibitem [{\citenamefont {Carroll~III}\ \emph {et~al.}(1994)\citenamefont
  {Carroll~III}, \citenamefont {Koepke}, \citenamefont {Amatucci},
  \citenamefont {Sheridan},\ and\ \citenamefont {Alport}}]{Caroll1994}%
  \BibitemOpen
  \bibfield  {author} {\bibinfo {author} {\bibfnamefont {J.~J.}\ \bibnamefont
  {Carroll~III}}, \bibinfo {author} {\bibfnamefont {M.~E.}\ \bibnamefont
  {Koepke}}, \bibinfo {author} {\bibfnamefont {W.~E.}\ \bibnamefont
  {Amatucci}}, \bibinfo {author} {\bibfnamefont {T.~E.}\ \bibnamefont
  {Sheridan}}, \ and\ \bibinfo {author} {\bibfnamefont {M.~J.}\ \bibnamefont
  {Alport}},\ }\href {\doibase 10.1063/1.1144590} {\bibfield  {journal}
  {\bibinfo  {journal} {Rev. Sci. Instrum.}\ }\textbf {\bibinfo {volume}
  {65}},\ \bibinfo {pages} {2991} (\bibinfo {year} {1994})}\BibitemShut
  {NoStop}%
\bibitem [{\citenamefont {Liziakin}\ \emph {et~al.}(2016)\citenamefont
  {Liziakin}, \citenamefont {Gavrikov}, \citenamefont {Murzaev}, \citenamefont
  {Usmanov},\ and\ \citenamefont {Smirnov}}]{Liziakin2016}%
  \BibitemOpen
  \bibfield  {author} {\bibinfo {author} {\bibfnamefont {G.~D.}\ \bibnamefont
  {Liziakin}}, \bibinfo {author} {\bibfnamefont {A.~V.}\ \bibnamefont
  {Gavrikov}}, \bibinfo {author} {\bibfnamefont {Y.~A.}\ \bibnamefont
  {Murzaev}}, \bibinfo {author} {\bibfnamefont {R.~A.}\ \bibnamefont
  {Usmanov}}, \ and\ \bibinfo {author} {\bibfnamefont {V.~P.}\ \bibnamefont
  {Smirnov}},\ }\bibfield  {booktitle} {\emph {\bibinfo {booktitle} {Physics of
  Plasmas}},\ }\href {\doibase 10.1063/1.4969084} {\bibfield  {journal}
  {\bibinfo  {journal} {Phys. Plasmas}\ }\textbf {\bibinfo {volume} {23}},\
  \bibinfo {pages} {123502} (\bibinfo {year} {2016})}\BibitemShut {NoStop}%
\bibitem [{\citenamefont {Liziakin}\ \emph {et~al.}(2017)\citenamefont
  {Liziakin}, \citenamefont {Gavrikov}, \citenamefont {Usmanov}, \citenamefont
  {Timirkhanov},\ and\ \citenamefont {Smirnov}}]{Liziakin2017}%
  \BibitemOpen
  \bibfield  {author} {\bibinfo {author} {\bibfnamefont {G.}~\bibnamefont
  {Liziakin}}, \bibinfo {author} {\bibfnamefont {A.}~\bibnamefont {Gavrikov}},
  \bibinfo {author} {\bibfnamefont {R.}~\bibnamefont {Usmanov}}, \bibinfo
  {author} {\bibfnamefont {R.}~\bibnamefont {Timirkhanov}}, \ and\ \bibinfo
  {author} {\bibfnamefont {V.}~\bibnamefont {Smirnov}},\ }\href {\doibase
  10.1063/1.4998806} {\bibfield  {journal} {\bibinfo  {journal} {{AIP} Adv.}\
  }\textbf {\bibinfo {volume} {7}},\ \bibinfo {pages} {125108} (\bibinfo {year}
  {2017})}\BibitemShut {NoStop}%
\bibitem [{\citenamefont {Tuszewski}\ \emph {et~al.}(2012)\citenamefont
  {Tuszewski}, \citenamefont {Smirnov}, \citenamefont {Thompson}, \citenamefont
  {Korepanov}, \citenamefont {Akhmetov}, \citenamefont {Ivanov}, \citenamefont
  {Voskoboynikov}, \citenamefont {Schmitz}, \citenamefont {Barnes},
  \citenamefont {Binderbauer}, \citenamefont {Brown}, \citenamefont {Bui},
  \citenamefont {Clary}, \citenamefont {Conroy}, \citenamefont {Deng},
  \citenamefont {Dettrick}, \citenamefont {Douglass}, \citenamefont {Garate},
  \citenamefont {Glass}, \citenamefont {Gota}, \citenamefont {Guo},
  \citenamefont {Gupta}, \citenamefont {Gupta}, \citenamefont {Kinley},
  \citenamefont {Knapp}, \citenamefont {Longman}, \citenamefont {Hollins},
  \citenamefont {Li}, \citenamefont {Luo}, \citenamefont {Mendoza},
  \citenamefont {Mok}, \citenamefont {Necas}, \citenamefont {Primavera},
  \citenamefont {Ruskov}, \citenamefont {Schroeder}, \citenamefont {Sevier},
  \citenamefont {Sibley}, \citenamefont {Song}, \citenamefont {Sun},
  \citenamefont {Trask}, \citenamefont {Drie}, \citenamefont {Walters},\ and\
  \citenamefont {Wyman}}]{Tuszewski2012}%
  \BibitemOpen
  \bibfield  {author} {\bibinfo {author} {\bibfnamefont {M.}~\bibnamefont
  {Tuszewski}}, \bibinfo {author} {\bibfnamefont {A.}~\bibnamefont {Smirnov}},
  \bibinfo {author} {\bibfnamefont {M.~C.}\ \bibnamefont {Thompson}}, \bibinfo
  {author} {\bibfnamefont {S.}~\bibnamefont {Korepanov}}, \bibinfo {author}
  {\bibfnamefont {T.}~\bibnamefont {Akhmetov}}, \bibinfo {author}
  {\bibfnamefont {A.}~\bibnamefont {Ivanov}}, \bibinfo {author} {\bibfnamefont
  {R.}~\bibnamefont {Voskoboynikov}}, \bibinfo {author} {\bibfnamefont
  {L.}~\bibnamefont {Schmitz}}, \bibinfo {author} {\bibfnamefont
  {D.}~\bibnamefont {Barnes}}, \bibinfo {author} {\bibfnamefont {M.~W.}\
  \bibnamefont {Binderbauer}}, \bibinfo {author} {\bibfnamefont
  {R.}~\bibnamefont {Brown}}, \bibinfo {author} {\bibfnamefont {D.~Q.}\
  \bibnamefont {Bui}}, \bibinfo {author} {\bibfnamefont {R.}~\bibnamefont
  {Clary}}, \bibinfo {author} {\bibfnamefont {K.~D.}\ \bibnamefont {Conroy}},
  \bibinfo {author} {\bibfnamefont {B.~H.}\ \bibnamefont {Deng}}, \bibinfo
  {author} {\bibfnamefont {S.~A.}\ \bibnamefont {Dettrick}}, \bibinfo {author}
  {\bibfnamefont {J.~D.}\ \bibnamefont {Douglass}}, \bibinfo {author}
  {\bibfnamefont {E.}~\bibnamefont {Garate}}, \bibinfo {author} {\bibfnamefont
  {F.~J.}\ \bibnamefont {Glass}}, \bibinfo {author} {\bibfnamefont
  {H.}~\bibnamefont {Gota}}, \bibinfo {author} {\bibfnamefont {H.~Y.}\
  \bibnamefont {Guo}}, \bibinfo {author} {\bibfnamefont {D.}~\bibnamefont
  {Gupta}}, \bibinfo {author} {\bibfnamefont {S.}~\bibnamefont {Gupta}},
  \bibinfo {author} {\bibfnamefont {J.~S.}\ \bibnamefont {Kinley}}, \bibinfo
  {author} {\bibfnamefont {K.}~\bibnamefont {Knapp}}, \bibinfo {author}
  {\bibfnamefont {A.}~\bibnamefont {Longman}}, \bibinfo {author} {\bibfnamefont
  {M.}~\bibnamefont {Hollins}}, \bibinfo {author} {\bibfnamefont {X.~L.}\
  \bibnamefont {Li}}, \bibinfo {author} {\bibfnamefont {Y.}~\bibnamefont
  {Luo}}, \bibinfo {author} {\bibfnamefont {R.}~\bibnamefont {Mendoza}},
  \bibinfo {author} {\bibfnamefont {Y.}~\bibnamefont {Mok}}, \bibinfo {author}
  {\bibfnamefont {A.}~\bibnamefont {Necas}}, \bibinfo {author} {\bibfnamefont
  {S.}~\bibnamefont {Primavera}}, \bibinfo {author} {\bibfnamefont
  {E.}~\bibnamefont {Ruskov}}, \bibinfo {author} {\bibfnamefont {J.~H.}\
  \bibnamefont {Schroeder}}, \bibinfo {author} {\bibfnamefont {L.}~\bibnamefont
  {Sevier}}, \bibinfo {author} {\bibfnamefont {A.}~\bibnamefont {Sibley}},
  \bibinfo {author} {\bibfnamefont {Y.}~\bibnamefont {Song}}, \bibinfo {author}
  {\bibfnamefont {X.}~\bibnamefont {Sun}}, \bibinfo {author} {\bibfnamefont
  {E.}~\bibnamefont {Trask}}, \bibinfo {author} {\bibfnamefont {A.~D.~V.}\
  \bibnamefont {Drie}}, \bibinfo {author} {\bibfnamefont {J.~K.}\ \bibnamefont
  {Walters}}, \ and\ \bibinfo {author} {\bibfnamefont {M.~D.}\ \bibnamefont
  {Wyman}},\ }\href {\doibase 10.1103/physrevlett.108.255008} {\bibfield
  {journal} {\bibinfo  {journal} {Phys. Rev. Lett.}\ }\textbf {\bibinfo
  {volume} {108}},\ \bibinfo {pages} {255008} (\bibinfo {year}
  {2012})}\BibitemShut {NoStop}%
\bibitem [{\citenamefont {Gilmore}\ \emph {et~al.}(2014)\citenamefont
  {Gilmore}, \citenamefont {Lynn}, \citenamefont {Desjardins}, \citenamefont
  {Zhang}, \citenamefont {Watts}, \citenamefont {Hsu}, \citenamefont {Betts},
  \citenamefont {Kelly},\ and\ \citenamefont {Schamiloglu}}]{Gilmore2014}%
  \BibitemOpen
  \bibfield  {author} {\bibinfo {author} {\bibfnamefont {M.}~\bibnamefont
  {Gilmore}}, \bibinfo {author} {\bibfnamefont {A.~G.}\ \bibnamefont {Lynn}},
  \bibinfo {author} {\bibfnamefont {T.~R.}\ \bibnamefont {Desjardins}},
  \bibinfo {author} {\bibfnamefont {Y.}~\bibnamefont {Zhang}}, \bibinfo
  {author} {\bibfnamefont {C.}~\bibnamefont {Watts}}, \bibinfo {author}
  {\bibfnamefont {S.~C.}\ \bibnamefont {Hsu}}, \bibinfo {author} {\bibfnamefont
  {S.}~\bibnamefont {Betts}}, \bibinfo {author} {\bibfnamefont
  {R.}~\bibnamefont {Kelly}}, \ and\ \bibinfo {author} {\bibfnamefont
  {E.}~\bibnamefont {Schamiloglu}},\ }\href {\doibase
  10.1017/s0022377814000919} {\bibfield  {journal} {\bibinfo  {journal} {J.
  Plasma Phys.}\ }\textbf {\bibinfo {volume} {81}},\ \bibinfo {pages}
  {345810104} (\bibinfo {year} {2014})}\BibitemShut {NoStop}%
\bibitem [{\citenamefont {Gilmore}\ \emph {et~al.}(2009)\citenamefont
  {Gilmore}, \citenamefont {Xie}, \citenamefont {Yan}, \citenamefont {Watts},\
  and\ \citenamefont {Lynn}}]{Gilmore2009}%
  \BibitemOpen
  \bibfield  {author} {\bibinfo {author} {\bibfnamefont {M.}~\bibnamefont
  {Gilmore}}, \bibinfo {author} {\bibfnamefont {S.}~\bibnamefont {Xie}},
  \bibinfo {author} {\bibfnamefont {L.}~\bibnamefont {Yan}}, \bibinfo {author}
  {\bibfnamefont {C.}~\bibnamefont {Watts}}, \ and\ \bibinfo {author}
  {\bibfnamefont {A.}~\bibnamefont {Lynn}},\ }in\ \href@noop {} {\emph
  {\bibinfo {booktitle} {Proceedings of the 36th EPS Conference on Plasma
  Physics}}}\ (\bibinfo {year} {2009})\BibitemShut {NoStop}%
\bibitem [{\citenamefont {Koepke}\ \emph {et~al.}(2008)\citenamefont {Koepke},
  \citenamefont {Finnegan}, \citenamefont {Vincena}, \citenamefont {Knudsen},\
  and\ \citenamefont {Chaston}}]{Koepke2008}%
  \BibitemOpen
  \bibfield  {author} {\bibinfo {author} {\bibfnamefont {M.~E.}\ \bibnamefont
  {Koepke}}, \bibinfo {author} {\bibfnamefont {S.~M.}\ \bibnamefont
  {Finnegan}}, \bibinfo {author} {\bibfnamefont {S.}~\bibnamefont {Vincena}},
  \bibinfo {author} {\bibfnamefont {D.~J.}\ \bibnamefont {Knudsen}}, \ and\
  \bibinfo {author} {\bibfnamefont {C.}~\bibnamefont {Chaston}},\ }\href
  {\doibase 10.1088/0741-3335/50/7/074004} {\bibfield  {journal} {\bibinfo
  {journal} {Plasma Phys. Controlled Fusion}\ }\textbf {\bibinfo {volume}
  {50}},\ \bibinfo {pages} {074004} (\bibinfo {year} {2008})}\BibitemShut
  {NoStop}%
\bibitem [{\citenamefont {Gueroult}\ \emph {et~al.}(2016)\citenamefont
  {Gueroult}, \citenamefont {Evans}, \citenamefont {Zweben}, \citenamefont
  {Fisch},\ and\ \citenamefont {Levinton}}]{Gueroult2016a}%
  \BibitemOpen
  \bibfield  {author} {\bibinfo {author} {\bibfnamefont {R.}~\bibnamefont
  {Gueroult}}, \bibinfo {author} {\bibfnamefont {E.~S.}\ \bibnamefont {Evans}},
  \bibinfo {author} {\bibfnamefont {S.~J.}\ \bibnamefont {Zweben}}, \bibinfo
  {author} {\bibfnamefont {N.~J.}\ \bibnamefont {Fisch}}, \ and\ \bibinfo
  {author} {\bibfnamefont {F.}~\bibnamefont {Levinton}},\ }\href {\doibase
  10.1088/0963-0252/25/3/035024} {\bibfield  {journal} {\bibinfo  {journal}
  {Plasma Sources Sci. Technol.}\ }\textbf {\bibinfo {volume} {25}},\ \bibinfo
  {pages} {035024} (\bibinfo {year} {2016})}\BibitemShut {NoStop}%
\bibitem [{\citenamefont {Fetterman}\ and\ \citenamefont
  {Fisch}(2008)}]{Fetterman2008}%
  \BibitemOpen
  \bibfield  {author} {\bibinfo {author} {\bibfnamefont {A.~J.}\ \bibnamefont
  {Fetterman}}\ and\ \bibinfo {author} {\bibfnamefont {N.~J.}\ \bibnamefont
  {Fisch}},\ }\href {\doibase 10.1103/PhysRevLett.101.205003} {\bibfield
  {journal} {\bibinfo  {journal} {Phys. Rev. Lett.}\ }\textbf {\bibinfo
  {volume} {101}},\ \bibinfo {pages} {205003} (\bibinfo {year}
  {2008})}\BibitemShut {NoStop}%
\bibitem [{\citenamefont {Fisch}\ and\ \citenamefont
  {Rax}(1992{\natexlab{a}})}]{Fisch1992}%
  \BibitemOpen
  \bibfield  {author} {\bibinfo {author} {\bibfnamefont {N.~J.}\ \bibnamefont
  {Fisch}}\ and\ \bibinfo {author} {\bibfnamefont {J.-M.}\ \bibnamefont
  {Rax}},\ }\href {\doibase 10.1103/PhysRevLett.69.612} {\bibfield  {journal}
  {\bibinfo  {journal} {Phys. Rev. Lett.}\ }\textbf {\bibinfo {volume} {69}},\
  \bibinfo {pages} {612} (\bibinfo {year} {1992}{\natexlab{a}})}\BibitemShut
  {NoStop}%
\bibitem [{\citenamefont {Fetterman}\ and\ \citenamefont
  {Fisch}(2009)}]{Fetterman2009}%
  \BibitemOpen
  \bibfield  {author} {\bibinfo {author} {\bibfnamefont {A.~J.}\ \bibnamefont
  {Fetterman}}\ and\ \bibinfo {author} {\bibfnamefont {N.~J.}\ \bibnamefont
  {Fisch}},\ }\href {\doibase 10.1088/0963-0252/18/4/045003} {\bibfield
  {journal} {\bibinfo  {journal} {Plasma Sources Sci. Technol.}\ }\textbf
  {\bibinfo {volume} {18}},\ \bibinfo {pages} {045003} (\bibinfo {year}
  {2009})}\BibitemShut {NoStop}%
\bibitem [{\citenamefont {Fisch}\ and\ \citenamefont
  {Rax}(1992{\natexlab{b}})}]{Fisch1992a}%
  \BibitemOpen
  \bibfield  {author} {\bibinfo {author} {\bibfnamefont {N.~J.}\ \bibnamefont
  {Fisch}}\ and\ \bibinfo {author} {\bibfnamefont {J.-M.}\ \bibnamefont
  {Rax}},\ }\href {\doibase 10.1088/0029-5515/32/4/i02} {\bibfield  {journal}
  {\bibinfo  {journal} {Nucl. Fusion}\ }\textbf {\bibinfo {volume} {32}},\
  \bibinfo {pages} {549} (\bibinfo {year} {1992}{\natexlab{b}})}\BibitemShut
  {NoStop}%
\bibitem [{\citenamefont {Rax}, \citenamefont {Gueroult},\ and\ \citenamefont
  {Fisch}(2017)}]{Rax2017}%
  \BibitemOpen
  \bibfield  {author} {\bibinfo {author} {\bibfnamefont {J.~M.}\ \bibnamefont
  {Rax}}, \bibinfo {author} {\bibfnamefont {R.}~\bibnamefont {Gueroult}}, \
  and\ \bibinfo {author} {\bibfnamefont {N.~J.}\ \bibnamefont {Fisch}},\ }\href
  {\doibase 10.1063/1.4977919} {\bibfield  {journal} {\bibinfo  {journal}
  {Phys. Plasmas}\ }\textbf {\bibinfo {volume} {24}},\ \bibinfo {pages}
  {032504} (\bibinfo {year} {2017})}\BibitemShut {NoStop}%
\bibitem [{\citenamefont {Ochs}\ and\ \citenamefont {Fisch}(2017)}]{Ochs2017b}%
  \BibitemOpen
  \bibfield  {author} {\bibinfo {author} {\bibfnamefont {I.~E.}\ \bibnamefont
  {Ochs}}\ and\ \bibinfo {author} {\bibfnamefont {N.~J.}\ \bibnamefont
  {Fisch}},\ }\href {\doibase 10.1063/1.4991510} {\bibfield  {journal}
  {\bibinfo  {journal} {Phys. Plasmas}\ }\textbf {\bibinfo {volume} {24}},\
  \bibinfo {pages} {092513} (\bibinfo {year} {2017})}\BibitemShut {NoStop}%
\end{thebibliography}

%

\end{document}